\documentclass[12pt]{article}
\usepackage[pdftex]{graphicx}
\usepackage{amssymb}

\title{A new universal cosmic-ray knee near the magnetic rigidity 10~TV with the NUCLEON space observatory}

\author{
E.\,Atkin$^+$,
V.\,Bulatov$^\dag$,
V.\,Dorokhov$^\dag$,
N.\,Gorbunov$^{\#\&}$,
S.\,Filippov$^\dag$,\\
V.\,Grebenyuk$^{\#\&}$,
D.\,Karmanov$^\times$,
I.\,Kovalev$^\times$,
I.\,Kudryashov$^\times$,\\
A.\,Kurganov$^\times$,
M.\,Merkin$^\times$,
A.\,Panov$^\times$\thanks{e-mail: panov@dec1.sinp.msu.ru},
D.\,Podorozhny$^\times$,
D.\,Polkov$^\dag$,\\
S.\,Porokhovoy$^\#$,
V.\,Shumikhin$^+$,
A.\,Tkachenko$^{\#\S}$,
L.\,Tkachev$^{\#\&}$,\\
A.\,Turundaevskiy$^\times$,
O.\,Vasiliev$^\times$,
A.\,Voronin$^\times$}

\date{}

\begin{document}

\maketitle

{
$^+$ National Research Nuclear University “MEPhI”, Kashirskoe Highway, 31. Moscow, 115409, Russia\\
$^\dag$ GORIZONT, LLC, Mamin-Sibiryak str, 145, Ekaterinburg, 620075, Russia\\
$^\#$ Joint Institute for Nuclear Research, Dubna, Joliot-Curie, 6, Moscow Region, 141980, Russia\\
$^\&$ “DUBNA” University, Universitetskaya str., 19, Dubna, Moscow region, 141980, Russia\\
$^\times$ Skobeltsyn Institute of Nuclear Physics, Moscow State University, 1(2), Leninskie Gory, GSP-1, Moscow, 119991, Russia\\
$^\S$ Bogolyubov Institute for Theoretical Physics, 14-b Metrolohichna Str., Kiev, 03143, Ukraine
}

\abstract{\it Data from the NUCLEON space observatory give a strong indication of the existence of a new universal cosmic ray ``knee'', which is observed in all groups of nuclei, including heavy nuclei, near a magnetic rigidity of about 10\,TV. Universality means the same position of the knee in the magnetic rigidity scale for all groups of nuclei. The knee is observed by both methods of measurement of particles energy implemented in the NUCLEON observatory---the calorimetric method and the kinematic method KLEM. This new cosmic ray ``knee'' is probably connected with the limit of acceleration of cosmic rays by some generic or nearby source of cosmic rays.}

\vspace{0.5cm}

The main mechanism of the acceleration of galactic cosmic rays with energy above approximately 1~GeV per nucleon is thought to be the acceleration of charged particles by magnetic fields of the termination shocks of supernova remnants \cite{GINZBURG1964,CRPROP-PTUSKIN1975-ENG,CRA-BLANDFORD1980-ApJ,CRA-BLANDFORD1987-PhysRep,CRA-AXFORD1981-ICRC,GAISSER1990,CRA-PTUSKIN2007}. The simplest theories of the acceleration of cosmic rays by termination shocks predict a simple power-law energy source spectrum of nuclei of cosmic rays with universal spectral index close to 2.0 \cite{CRA-KRYMSKY1977,CRA-BELL1978-I}. Consequently, the observed spectra are all expected to be power-laws with about the same values of spectral indices. However, recent direct balloon and space experiments have shown that the real situation is much more complicated. First, it was discovered that the protons and helium energy spectra have different spectral indices \cite{ATIC-2004-ZATSEPIN-IzvRan,ATIC-2009-PANOV-IzvRAN-ENG,CREAM2009B,CREAM2011-ApJ-PHe-I,CR-PAMELA-2011-p-He-Mag,PAMELA-2011-p-He-Cal,AMS-02-2015-PRL-p,AMS-02-2015-PRL-He}. Then, it was discovered that the spectra deviate significantly from a single power-law form at energies even lower than the energy of the famous cosmic-ray knee (about $3\cdot10^{15}$\,eV). For example, a nearly universal hardening of the spectra near the magnetic rigidities at about 200--500\,GV was observed in the spectra of all abundant cosmic-ray nuclei \cite{ATIC-2009-PANOV-IzvRAN-ENG,CR-CREAM2010A,CR-PAMELA-2011-p-He-Mag,AMS-02-2015-PRL-p,AMS-02-2015-PRL-He}. In addition, there are a number of indications to other features in the energy spectra of cosmic rays. 
\begin{figure}
\centering
\includegraphics[width=0.66\textwidth]{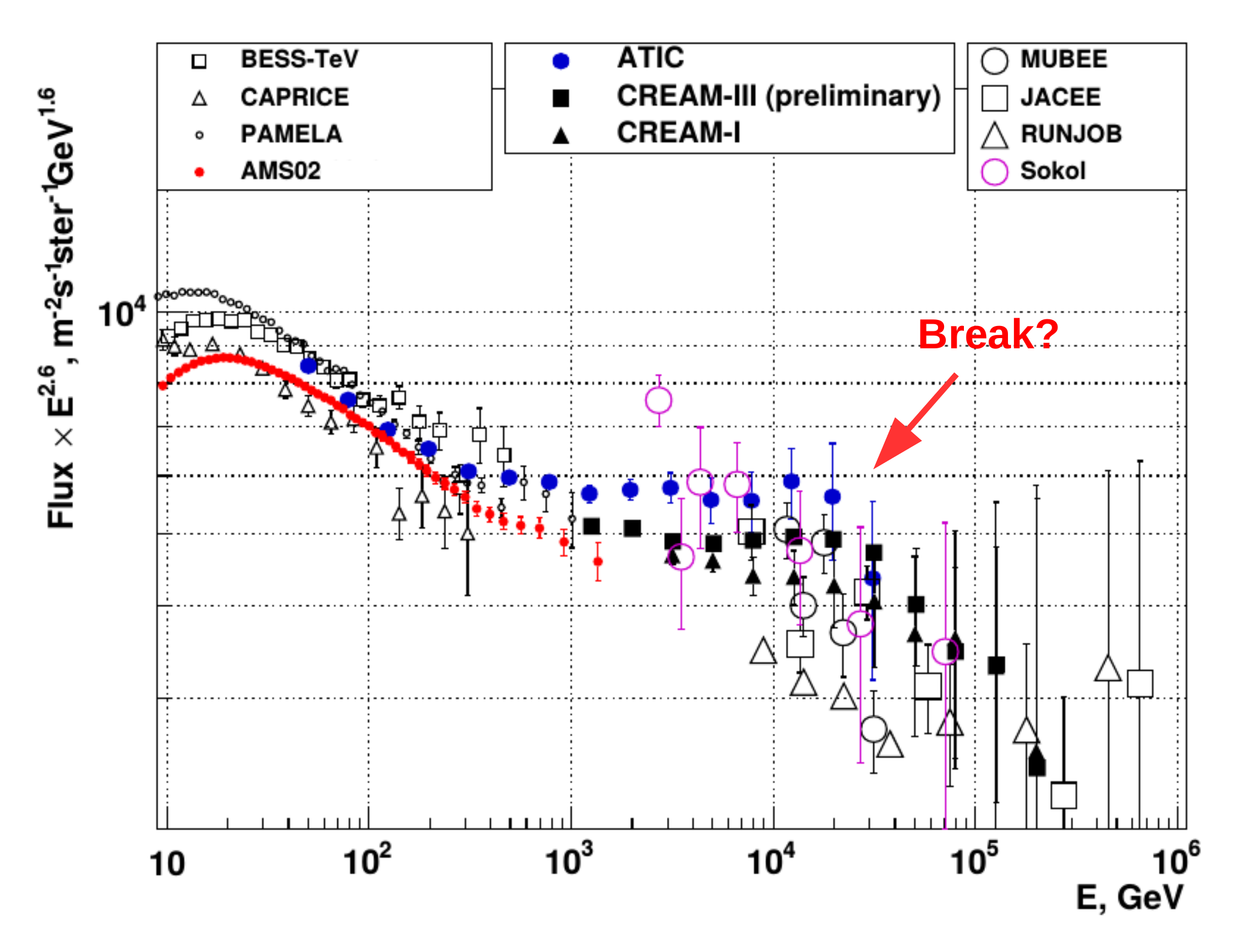}
\caption{\label{fig:ProtonsCompilation} A compilation of the data on proton spectrum before the NUCLEON experiment and very recent data from the CREAM experiment \cite{CR-CREAM2017-ApJ-pHe}. The shown spectra are: AMS02 \cite{AMS-02-2015-PRL-p}, BESS-TeV \cite{BESS-TeV-2003,BESS-TeV-2004,BESS-TeV-2005}, CAPRICE \cite{CAPRICE-2003}, PAMELA \cite{CR-PAMELA-2011-p-He-Mag}, ATIC \cite{ATIC-2009-PANOV-IzvRAN-ENG}, CREAM-III (preliminary) \cite{CREAM2009B}, CREAM-I \cite{CREAM2011-ApJ-PHe-I}, MUBEE \cite{MUBEE-1993-JetpLett,MUBEE-1994-YadFiz}, JACEE \cite{JACEE-1998-ApJ}, RUNJOB \cite{RUNJOB-2005-ApJ}, SOKOL \cite{SOKOL-1993-IzvRAN}.}
\end{figure}
For example, one can see an indication of a break near 10\,TeV in the proton spectrum in the collection of data from different experiments (Fig.~\ref{fig:ProtonsCompilation}); however, no experiment has been able to provide a statistically significant result. This break was discussed specifically in two recent papers on the NUCLEON experiment \cite{NUCLEON-2017-JCAP} and CREAM experiment \cite{CR-CREAM2017-ApJ-pHe}. It was pointed out in both papers that there is also an indication of a break near the same magnetic rigidity in the helium spectrum 10\,TV but estimations of the statistical significance of the breaks were not presented in either paper. Therefore, it is still important to prove that the spectral break near the magnetic rigidity of 10\,TV really exists in the spectra of protons and helium nuclei with sufficiently high statistical significance. 

The problem is actually even more interesting. V.~Zatsepin and N.~Sokolskaya, based on data like that in Fig.~\ref{fig:ProtonsCompilation}, in their paper 
\cite{CR-ZATSEPIN2006} suggested that the break in the spectra of protons and helium not only really exists but that it also has a universal nature in the sense that it takes place in the spectra of all nuclei near the same \emph{magnetic rigidity}. Since an acceleration limint is determined by the magentic rigidity of the particles, this break was associated with a certain type of cosmic ray source, which had an acceleration limit near 10\,TV. Using this hypothesis and some other facts and assumptions, they constructed a phenomenological three-component model of the spectra of cosmic rays, which in a certain approximation reasonably approximated the energy spectra of all cosmic-ray nuclei at all energies up to $10^{17}$\,eV. One can say that this model \emph{predicts} that a break near the magnetic rigidity of about 10\,TV should also be detected in rigidity spectra of primary heavy nuclei $Z\geq6$. However, this possibility has not yet been studied experimentally due to a lack of statistics from the previous direct cosmic-ray experiments in related rigidity regions. Hence, the task is not only to confirm the existence of a break in the spectra of protons and helium with sufficient statistical significance but also to detect similar break in the spectra of heavy primary nuclei. Exactly this group of problems is studied in the present letter with  data from the NUCLEON space spectrometer. NUCLEON is the first experiment that can provide sufficient statistics in the spectra of heavy nuclei for magnetic rigidity above 10\,TV.

\begin{figure}
\centering
\includegraphics[width=0.66\textwidth]{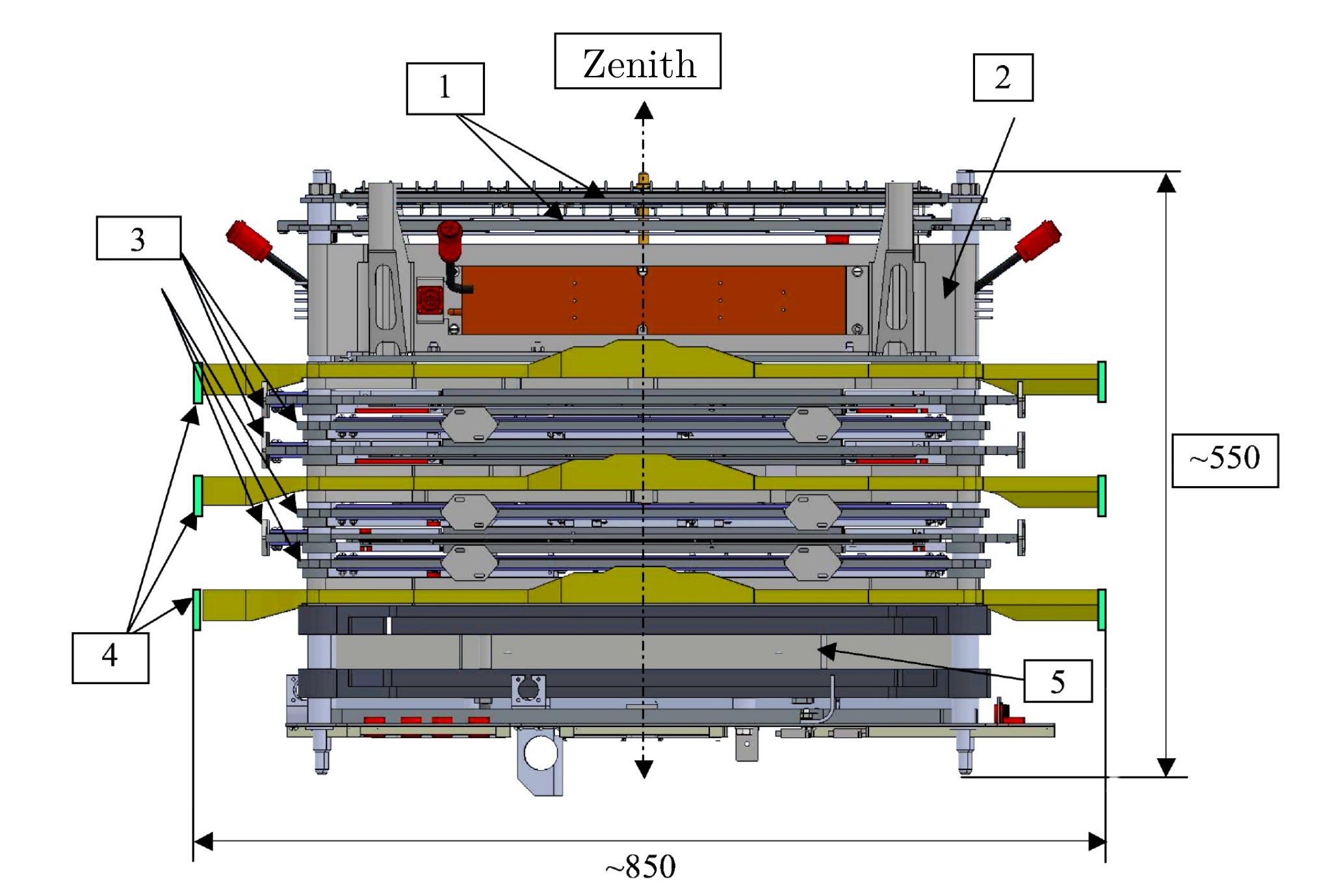}
\caption{The NUCLEON detector simplified scheme: 1. charge measurement system, 2. carbon target, 3. KLEM system tracker, 4. scintillator trigger system, and 5. ionization calorimeter (IC).}
\label{fig:NUCLEON}
\end{figure}

The NUCLEON space spectrometer was mainly designed to measure the spectra of cosmic ray nuclei with an individual charge resolution in an energy range from a few TeV to 1\,PeV per particle. Two different energy measurement methods are implemented in the NUCLEON apparatus: the first uses an ionization calorimeter (IC), and the second is a kinematic method---the Kinematic Lightweight Energy Meter (KLEM) \cite{KLEM-2002,KLEM-2005B}, which is based on the measurement of the multiplicity of secondary particles after the first nuclear interaction of a primary particle with a target in the spectrometer. The KLEM method is a principally new method of energy measurement that is used in practice for the first time. The main systems of the spectrometer (see Fig.~\ref{fig:NUCLEON}) are four planes of the charge measurement system, a carbon target, six planes of the energy measurement system using the KLEM method (KLEM system tracker), three double-layer planes of the scintillator trigger system, and a small aperture ionization calorimeter ($25\times25$\,cm$^2$). Details of the detector design are provided in the articles \cite{NUCLEON-DEZ-2007A,NUCLEON-DEZ-2007B,NUCLEON-DEZ-2007C,NUCLEON-DEZ-2010,NUCLEON-DEZ-2015}. On December 28, 2014, the NUCLEON detector was launched into a sun-synchronous orbit. The recent NUCLEON results obtained after first two years of data acquisition are presented in the report \cite{NUCLEON-2017-ICRC-Highlite}. Details of the implementation of the methods of energy measurement with KLEM and calorimetric methods in the NUCLEON experiment are given in the paper \cite{NUCLEON-2017-JCAP}. In the present paper the analysis will be based on both the IC and the KLEM methods data.

\begin{figure*}
\centering
\includegraphics[width=0.49\textwidth]{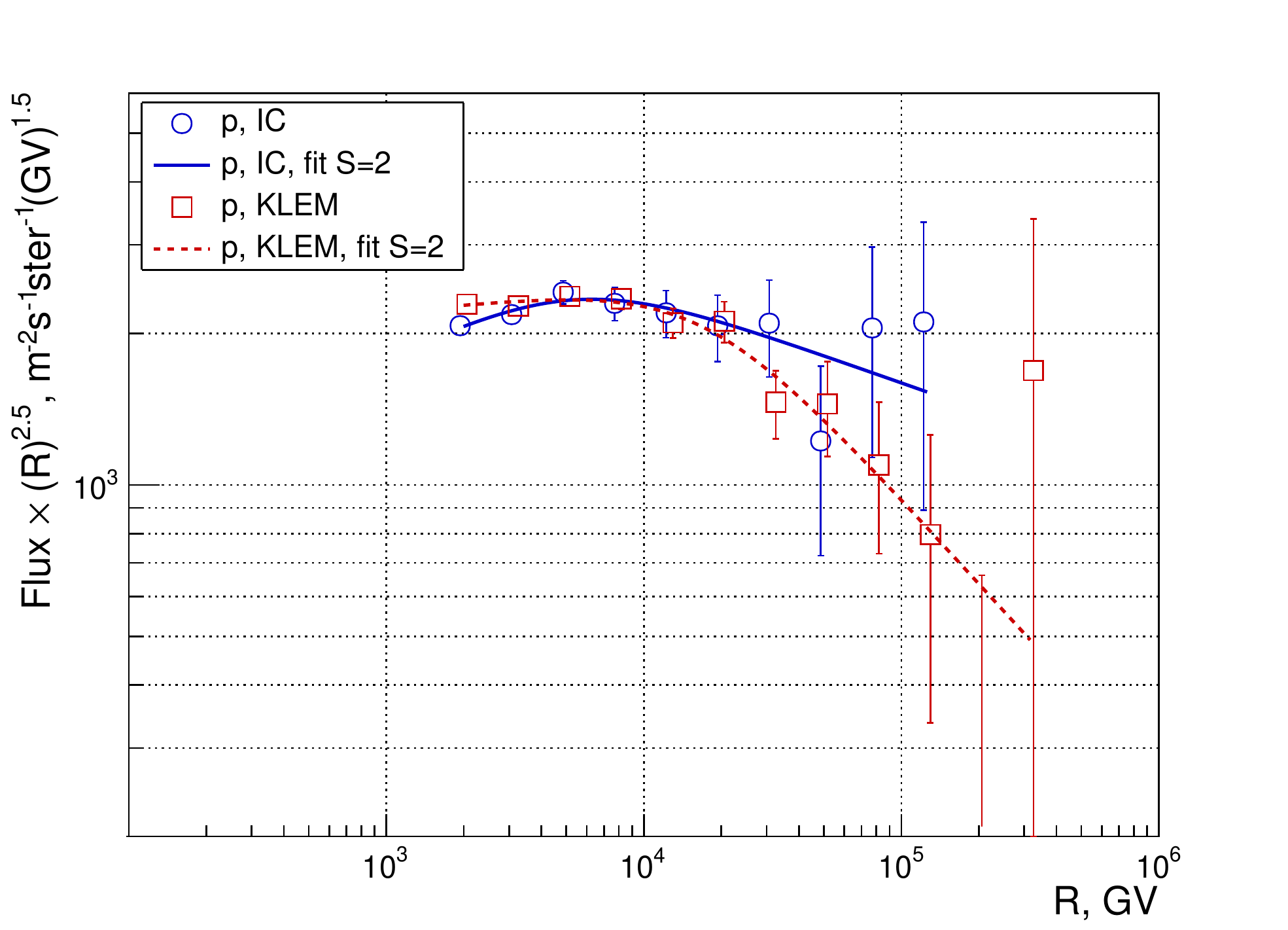}
\includegraphics[width=0.49\textwidth]{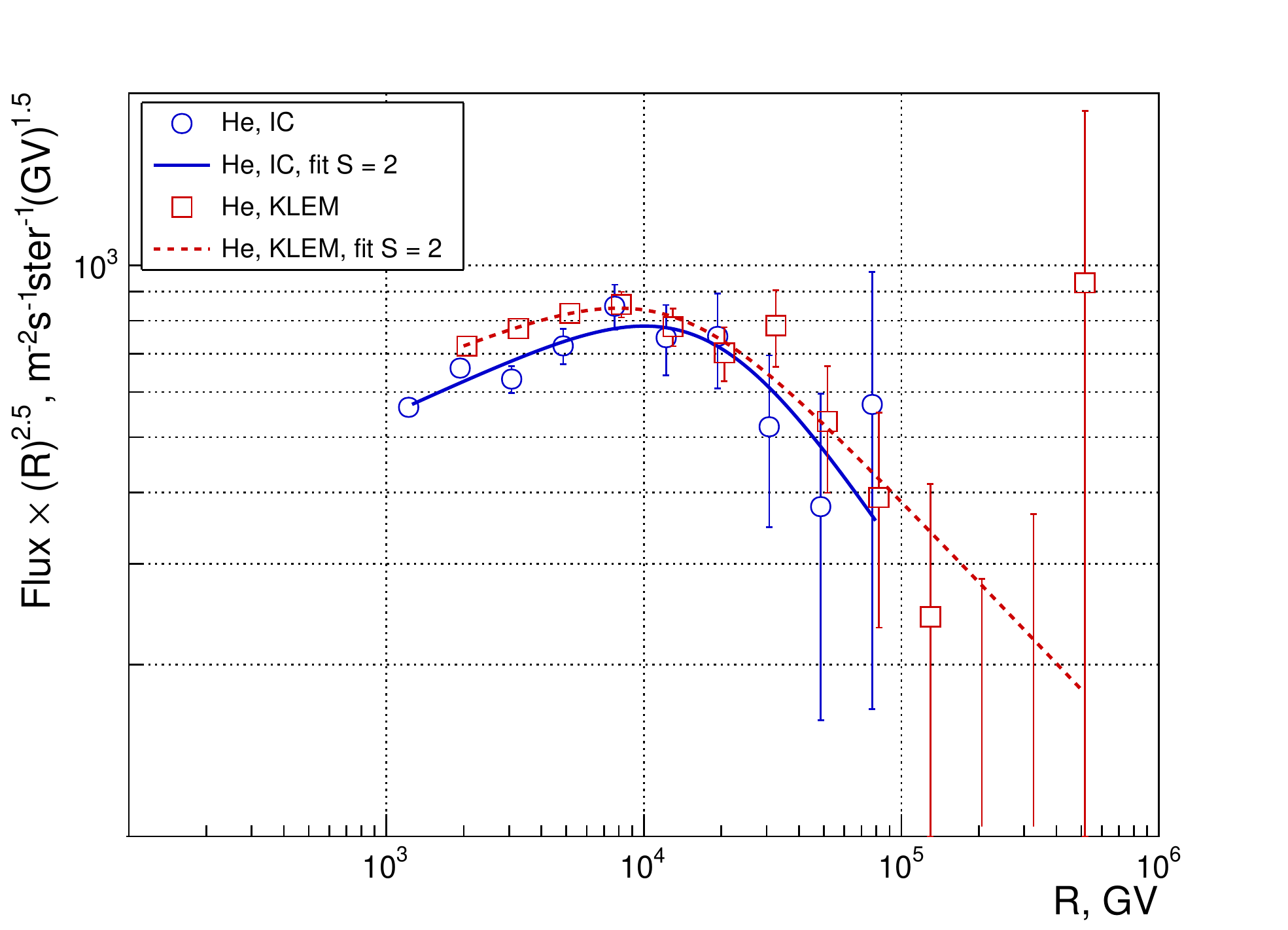}\\
\includegraphics[width=0.49\textwidth]{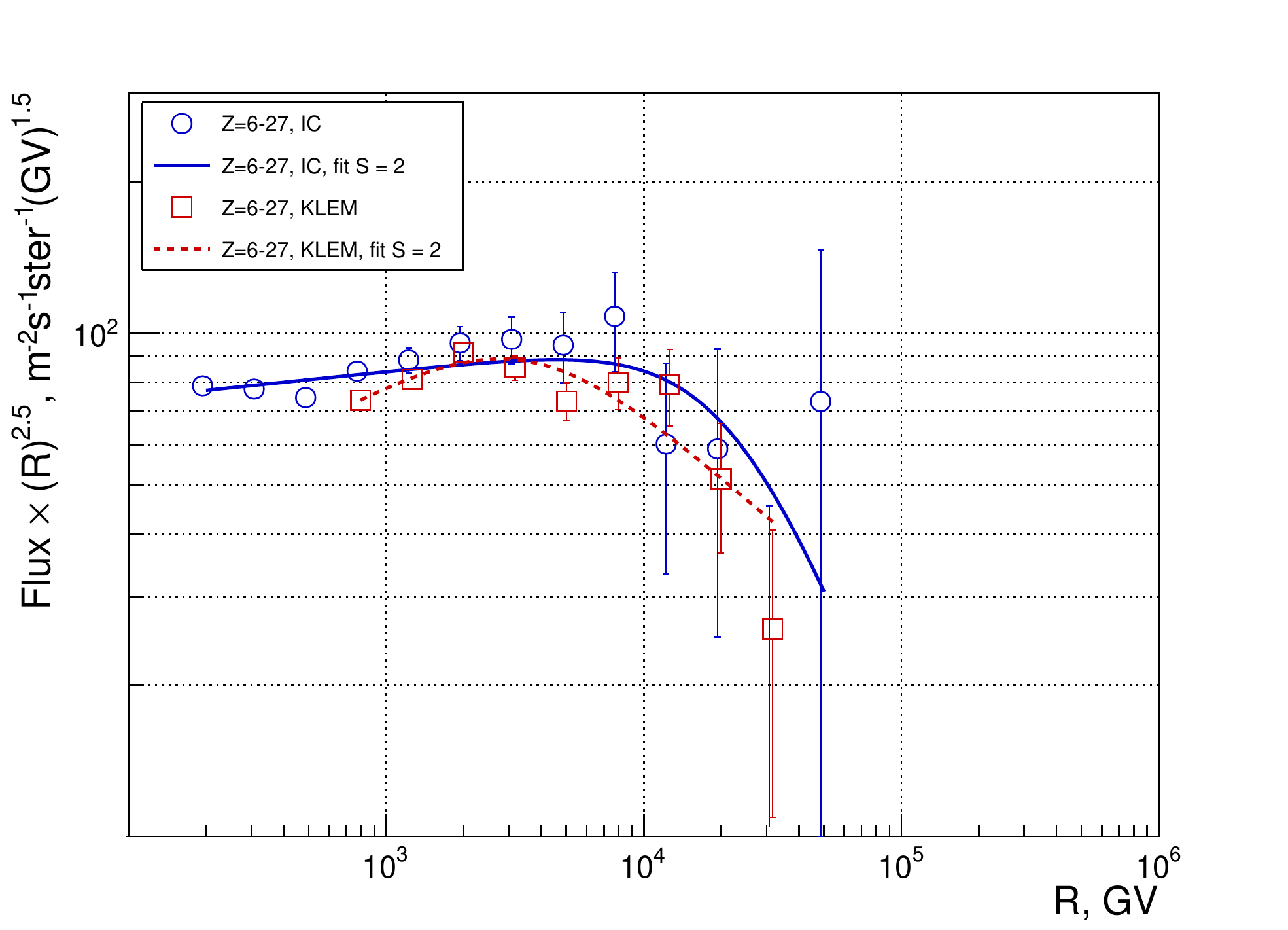}
\includegraphics[width=0.49\textwidth]{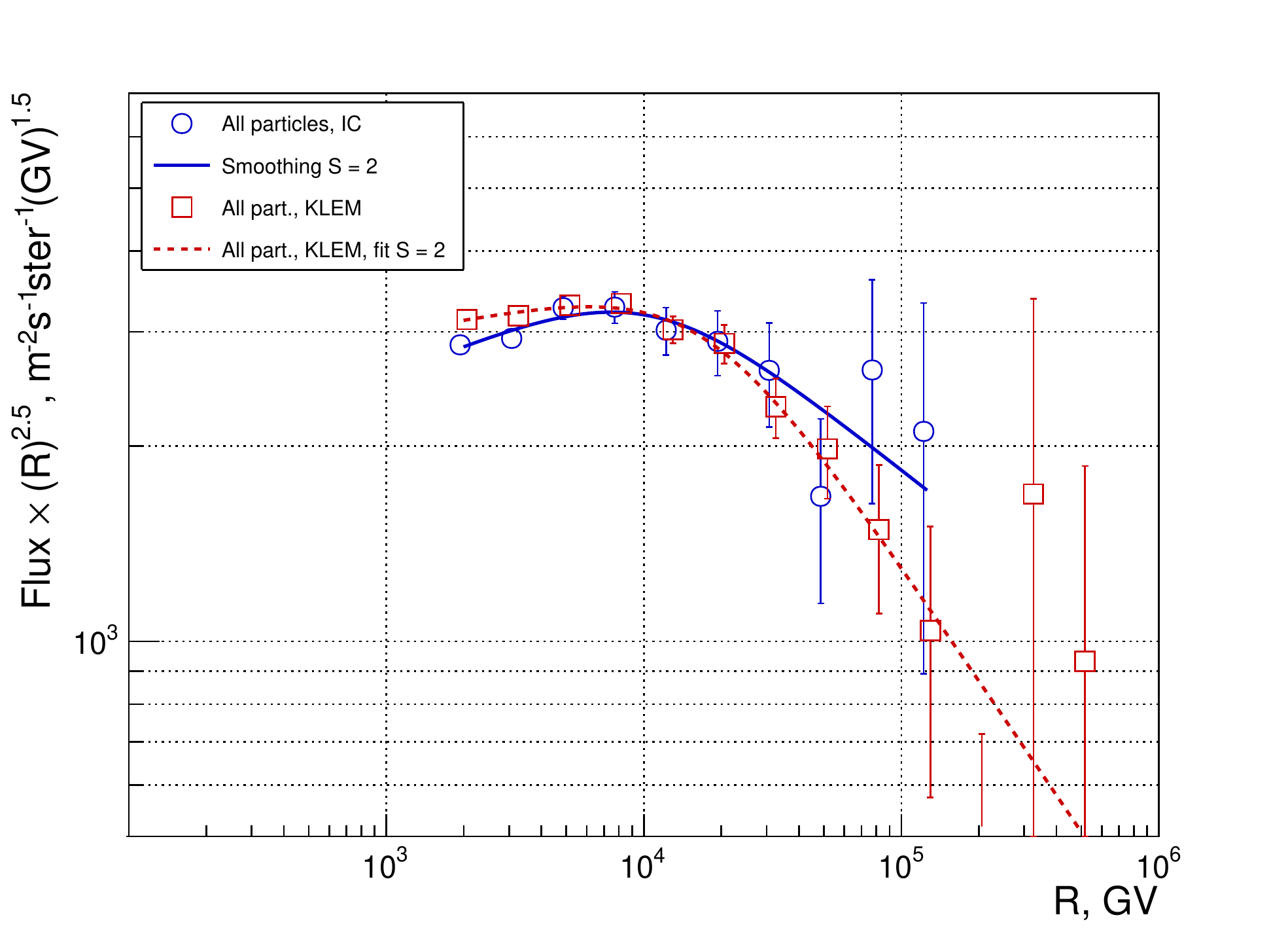}
\caption{\label{fig:DATA}The magnetic rigidity spectra for protons, for helium nuclei, for heavy nuclei with charges $Z=6 \div 27$ and for all particles measured by the calorimetric method (IC) and by the KLEM method in the NUCLEON experiment. The approximations of the spectra by the double-power-laws from the family of the functions (\ref{eq:ZSgen}) for the smoothing parameter $S=2$ are also shown. The spectra of IC are artificially shifted a bit to the left and spectra of KLEM are shifted to the right to avoid strong overlapping of the data points.}
\end{figure*}

The magnetic rigidity spectra of four groups of nuclei: protons, helium nuclei, joint heavy nuclei spectrum with charges $Z=6 \div 27$, and of all nuclei measured both by the calorimetric and KLEM methods in the NUCLEON experiment are shown in Fig.~\ref{fig:DATA} together with some sort of approximations of the spectra, which will be discussed later. The spectra of first three groups: p, He, $Z=6 \div 27$ are independent of each other. All nuclei spectrum depends on first three groups of course and considered here because it provides the best possible statistics.  All of the heavy nuclei $Z=6 \div 27$ were joint to one single rigidity spectrum because the statistics for each individual heavy nucleus is still too low for the analysis at the present stage of the experiment and, at the same time, the nuclei group $Z=6 \div 27$ contains all abundant heavy primary nuclei. Nuclei with charges from 3 to 5 did not considered specially since they are mainly secondary nuclei produced by spallation of primary heavier nuclei with interstellar gas during their propagation and the spectra of secondary nuclei are expected to be different from the spectra of primary nuclei: protons, helium, main part of nuclei $Z=6-27$. Visually, the presence of a break in each of the spectra is determined in the region between the 5\,TV and 20\,TV of magnetic rigidity. All of the studied groups of  spectra: p, He, $Z=6 \div 27$ and all nuclei show approximately similar spectral breaks near the rigidity of 10\,TV, which indicates the universal character of this ``knee.'' The existence of the break near 10\,TV in the spectra of \emph{heavy nuclei}, predicted by the three component model \cite{CR-ZATSEPIN2006} and has not been tested previously, is confirmed. This means that this break, as was supposed in \cite{CR-ZATSEPIN2006}, actually has a universal nature and may be related to an acceleration limit of some cosmic-ray source. The exact statistical meaning of the existence of this universal break will be discussed in the main part of the following paper.

The errors that are shown in Fig.~\ref{fig:DATA} are purely statistical. It is seen that some energy bins of the spectra are empty and it is acceptable for methods applied to analysis of the data in the present paper as described below. It is expected that the systematic errors of the data are mainly related to uncertainties of the event trigger operation and they are estimated to be approximately 25\% in absolute intensity and $\pm0.06$ in the common spectral index of the spectra. The systematics related to energy determination by the both methods were tested in CERN beam tests experiments \cite{NUCLEON-DEZ-2010} and expected to be $\sim$5\%. The IC method and the KLEM method of energy measurement are in fact two different experiments with different physics of generation of the trigger, jointed in one experimental setup. The difference of the data of these two methods gives a natural measure of the systematics errors of both methods and it is seen from Fig.~\ref{fig:DATA} that this difference is not large. The qualitative behavior of the spectra of both methods is essentially the same. It is important that all probable systematics in the shape of the spectra (breaks, upturns, peaks and dips) produced by apparatus effects like energy saturation of the detectors or other artifacts may be related 17 May 2018only to total kinetic energy of the particles since both methods---IC and KLEM---operated in terms of total energy per particle. Therefore systematic artifacts in the spectra may in principle take place (if any) at the same energy per particle for different kinds of nuclei, but not at the same magnetic rigidity as in Fig.~\ref{fig:DATA}. For example, the position of the breaks for protons and nuclei $Z=6-27$ take place near 10~TeV and 100~TeV in terms of energy per particle and they can not have the same systematic origin. Therefore it is very unlikely that the break near the same magnetic rigidity 10~TV in different groups of nuclei have systematic origin and we may exclude this possibility. We consider the following analysis of the shapes of the spectra to be stable against supposed systematic errors.

Our task now is to estimate the statistical significance of the existence of a break in each spectrum of Fig.~\ref{fig:DATA} and to estimate the parameters of the breaks. There are two kinds of uncertainties in this problem: the first is the statistical uncertainty and the second is the uncertainty in the methodology. We start the discussion with the first, purely statistical, problem.

Let us suppose that we have selected some family of functions $F(\mathbf{a}, R)$ to be a model of a broken power law and that we would like to approximate one of the experimental plots in Fig.~\ref{fig:DATA} by one of the functions. Here, $\mathbf{a}$ is the set of parameters that should be fitted to approximate the data points by a curve of the family and $R$ is the rigidity. The rigidities of particles to obtain the spectra in Fig.~\ref{fig:DATA} were reconstructed from energy deposition of the calorimeter event-by-event, therefore the statistics of each single bin are Poisson up to the known normalization factor. The feature of the problem is that the statistics at the high-rigidity end of the spectra are small (some energy bins are even empty). Therefore, one cannot use the chi-square family of methods to approximate the data. However, a maximum likelihood method, which is based directly on Poisson statistics of counts in the rigidity bins, is still possible. If $N_i$ is the number of counts in the $i$-th bin of the spectrum, then it is easy to obtain that the likelihood function reads
\begin{equation}
 \mathcal{P}(\mathbf{a}) = -\sum_i[N_i\ln F(\mathbf{a},R_i) - \ln N_i! - F(\mathbf{a},R_i)].
 \label{eq:P}
\end{equation}
It can be seen that the expression (\ref{eq:P}) is meaningful, even if some $N_i$ is equal to zero. To find the optimal approximation of the data, one should minimize the likelihood function $\mathcal{P}(\mathbf{a})$ by the vector of parameters $\mathbf{a}$. It should be emphasized that the proposed method should not be mixed with the well-known family of chi-square methods with small numbers of counts \cite[P.734]{PRESS-BOOK} which provides small approximate analytical corrections for the usual chi-square method in the cases when the bin statistics do not satisfy the condition $N_i \gtrsim 10$, but all $N_i > 0$. The proposed method is exact and works for any $N_i$ but it is not analytical, it demands numerical minimization of the likelihood function (\ref{eq:P}) by the vector of parameters $\mathbf{a}$.

Now we turn to the methodological part of the problem. To represent a broken power law we use the function that was used for the same goal in the three-component model of Zatsepin-Sokolskaya \cite{CR-ZATSEPIN2006} and was used in the paper \cite{CR-HORANDEL-2003} to represent the known cosmic-ray knee near $3\cdot10^{15}$\,eV. This function may be written in the following symmetric form:
\begin{equation}
 F_S(\gamma_1,\gamma_2, A_*, R_*; R) = 
 A_*\left(\!\frac{R}{R_*}\!\right)^{-\frac{\gamma_1}{2}}\!\!\!
        \left(\!\frac{R}{R_*}\!\right)^{-\frac{\gamma_2}{2}}\!
 \left[ \frac{(R_*/R)^{\frac{S}{2}} + (R/R_*)^{\frac{S}{2}}}{2} \right]^{\frac{\gamma_1-\gamma_2}{S}}
 \label{eq:ZSgen}
\end{equation}
Here the parameter $S$ represents a degree of smoothing of the linking point of two power laws with spectral indices $\gamma_1$ and $\gamma_2$. The statistics of the NUCLEON data (Fig.~\ref{fig:DATA}) do not permit us to reasonably restrict $S$ and, therefore, we have to estimate the uncertainty of all of the results relative to the uncertainty of $S$. The function (\ref{eq:ZSgen}) has four parameters $\gamma_1,\gamma_2, A_*, R_*$ to be fitted and one parameter $S$ that we consider to be undefined. $R_*$ is some effective position of the breaking point and $A_*$ is the amplitude of the curve in this point. The special case $S\to\infty$ represents point-like linking of two power laws with indices $\gamma_1$ and $\gamma_2$ in the point $R_*$. 

In Fig.~\ref{fig:DATA}, the approximations of the data by the functions (\ref{eq:ZSgen}) with the smoothing parameter $S = 2$ that were obtained by minimization of the likelihood function (\ref{eq:P}) by $\gamma_1,\gamma_2, A_*, R_*$ are shown. For values $S<2$, the approximation does not look like a broken power law and, therefore, $S=2$ is the reasonable minimum for the parameter $S$ to be considered. The notion of two separate $\gamma_1$ and $\gamma_2$ become in fact meaningless in the context of the present problem for $S<2$. But the values of $S$ from 2 up to $+\infty$ should be studied.
\begin{table}
\caption{\label{tab:Tab1}Parameters of the break obtained by approximation of the double power law with different smoothing parameter $S$ for various nuclear groups in IC and KLEM methods}
\centering
\begin{tabular}{|l|c|c|c|c|}
 \hline
 nuclear & $\Delta\gamma$  &   R*(TV)    &     $\Delta\gamma$  &   R*(TV) \\
 \cline{2-5}
 group   & \multicolumn{2}{|c|}{$S=2$} & \multicolumn{2}{|c|}{$S=\infty$} \\
 \hline
 \multicolumn{5}{|c|}{IC}\\
 \hline
 p     &  0.44  &  5.45   &  0.27   &  5.38 \\
 He    &  0.89  &  18.18  &  0.50   &  9.65 \\
 Nucl  &  1.39  &  22.93  &  0.72   &  7.94 \\
 All   &  0.48  &  10.42  &  0.31   &  7.28 \\
 \hline
 \multicolumn{5}{|c|}{KLEM}\\
 \hline
 p     &  0.61  &  17.05  &  0.37   &  9.94 \\
 He    &  0.65  &  12.90  &  0.42   &  9.04 \\
 Nucl  &  0.72  &   3.52  &  0.43   &  2.13 \\
 All   &  0.66  &  17.13  &  0.40   & 10.11 \\
 \hline
\end{tabular}
\end{table}

Tab.~\ref{tab:Tab1} 
gives a measure of ​​the methodological stability of the results of approximation against the uncertainty of the smoothing parameter $S$. In this table, for each  nuclear group that we used, the values ​​of $\Delta\gamma=\gamma_2-\gamma_1$ and $R_*$ obtained in the two extreme assumptions $S = 2$ and $S = \infty$ are compared for both methods IC and KLEM. From Tab.~\ref{tab:Tab1} 
it follows that the dependence of the quantities $\Delta\gamma$ and $R_*$ on the assumption of smoothing parameter $S$ exists. Moreover, it can be seen that $\Delta\gamma$ is generally less for $S=\infty$ than for $S=2$, but it will be seen below that this difference does not influence essentially to the statistical significance of the existence of the break. Note that the value $R_*$ is between 2.13\,TV and 22.7\,TV in all nuclear groups and all models, and the geometrical mean of all these values is 11.3\,TV. This corresponds to the notion of a universal break with a position near $\sim10$\,TV up to a multiplicative factor of the order of the unit.

Another factor of a methodological uncertainty is related to the binning of the magnetic rigidity to build the rigidity spectra. The binnings down to eighth times more fine than used in Fig.~\ref{fig:DATA} were used to study the systematic stability of the results against this problem. It was proven that this factor of the systematic uncertainty is small relative to the statistical uncertainties and is not important for the final results. The details will be presented in the forthcoming papers.

The next step of the analysis is determination of the statistical significance of the statement about the existence of a break. This problem also was solved for two extremal cases $S = 2$ and $S = \infty$ to account for the methodological uncertainty for the factor $S$. Because of the essentially nonlinear method of approximation of the spectra and the low statistics in the high-energy part of the spectra, the problem of obtaining the statistical estimates can only be solved with the Monte Carlo method. For each of the smoothing parameters $S$ and for each nuclear group, for the generation of model spectra in the Monte Carlo method, the mean values ​​for generating Poisson random values ​​at each point of the spectrum are values ​​that give the corresponding approximations of the spectra (see Fig.~\ref{fig:DATA}). For each Monte Carlo model spectrum, a complete data processing procedure is performed to find an approximation function. The statistical distributions of the fit parameters that we have obtained will help to answer the question of the statistical reliability of the determination of various break parameters.

The statistical errors of experimental values are usually ​ estimated in terms of standard deviations. Such estimates suggest that random deviations are distributed approximately according to the Gaussian law. However, in our case, the distributions that we have obtained, generally speaking, have nothing to do with the Gaussian distribution and, therefore, the error estimates in terms of standard deviations are meaningless. A complete and exhaustive answer to the question of the statistical reliability of the parameters being determined can only be provided by the statistical distributions themselves. The question of interpreting these distributions should be posed separately.

\begin{figure}
\centering
\includegraphics[width=0.75\textwidth]{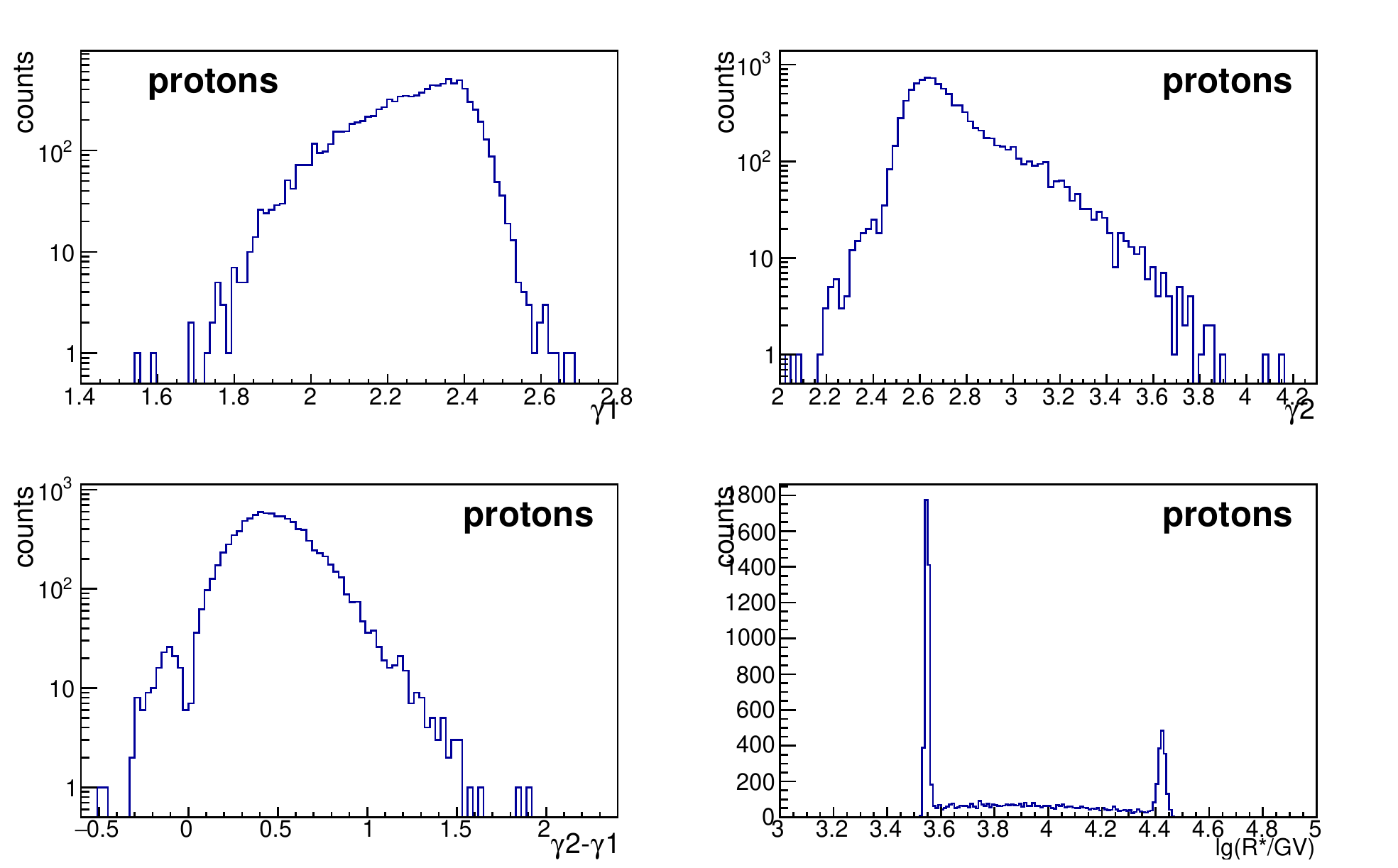}
\caption{\label{Fig:4S2} The distributions of the values $\gamma_1,\gamma_2,\gamma_2-\gamma_1,R_*$ in the spectrum of protons, for smoothing parameter $S=2$}
\end{figure}

The example of distributions of the values $\gamma_1,\gamma_2,\gamma_2-\gamma_1,R_*$ that we obtained in the Monte Carlo simulation for spectrum of protons in IC method for smoothing parameters $S=2$  are shown in Fig.~\ref{Fig:4S2}. The complete set of the pictures for all nuclear groups and for both values $S=2$ and $S=\infty$ is presented in the Supplement material to the paper \cite{SUPPLEMENT}.

The distribution functions for the parameter $R_*$ have a very bizarre shape. It should be noted that the positions of the narrow peaks at the edges of the distribution in either case do not correspond to the boundaries of the region used for the search for a minimum or to the boundary points of the spectra---they are always located in the inner region of the spectrum in terms of the magnetic rigidity. The unusual shape of the distribution is related to a very flat bottom of the likelihood function for the parameter $R_*$, limited at the same time by very steep walls. There is a high probability for the minimum to be near one of the ``bottom corner'' of such function---it is in fact a variant of realization of spontaneous symmetry breaking mechanisms.

\begin{table}
 \caption{\label{tab:Tab2}Probabilities for various conditions to the differences of the spectral indices at break \mbox{$\Delta\gamma = \gamma_2-\gamma_1$}.}
 \centering
\begin{tabular}{|l|c|c|c|}
\hline
nucl. &     $S=2$               & $S=\infty$         & single $\gamma$ \\
group &     percents($\sigma$)  & percents($\sigma$) & percents($\sigma$) \\
\hline
1 &     2  & 3 & 4 \\
\hline
\multicolumn{4}{|c|}{\strut IC}\\
\hline
p           & 99.31(2.70)      & 98.49(2.43)       & 53.17(0.73) \\        
He          & 99.85(3.25)      & 99.45(2.78)       & 90.13(1.65) \\
$6\div27$   & 95.94(2.05)      & 97.93(2.31)       & 90.00(1.64) \\
All         & 99.97(3.62)      & 99.89(3.26)       & 91.07(1.70) \\
\hline
\multicolumn{4}{|c|}{\strut KLEM}\\
\hline
p           & $>$99.99(3.9) & $>$99.99(3.9)  & 98.67(2.47)      \\
He          & $>$99.99(3.9) & $>$99.99(3.9)  & 99.59(2.93)      \\
$6\div27$   & $>$99.99(3.9) & $>$99.99(3.9)  & $>$99.99(3.9) \\
All         & $>$99.99(3.9) & $>$99.99(3.9)  & $>$99.99(3.9) \\
\hline
\end{tabular}
\end{table}

From the point of view of the possibility of the existence of a break in the spectrum, the most interesting value is the difference in the spectral indices after and before bending: $\Delta\gamma = \gamma_2-\gamma_1$. Formally, a break exists if this difference is greater than zero: $\gamma_2>\gamma_1$. Tab.~\ref{tab:Tab2}
in the columns 2 and 3 gives the statistical reliability in percents of the hypothesis $\Delta\gamma>0$ (the break exists) for both methods IC and KLEM, for smoothing parameters $S=2$ and $S=\infty$ and separately for each nuclear group. For convenience, the equivalent number of standard deviations of the normal distribution are given in parentheses. We, however, recall that the distributions of $\Delta\gamma$ really have little in common with the normal distribution, so the values in parentheses, strictly speaking, do not have a physical meaning. These data were obtained directly from the probability distributions like in Fig.~\ref{Fig:4S2}. 

In the IC method the probability that the break exists, $\Delta\gamma>0$, in some nuclear groups are rather high (formally greater than $3\sigma$) but in some others it is lower than $3\sigma$. The probabilities for $S=2$ and $S=\infty$ are comparable, therefore the systematics related to uncertainty of $S$ (Tab.~1) do not influence essentially the results related to the significance of the break.  The data statistics in the KLEM method is approximately twice larger than in the IC method and as a result the estimated statistical significance of the break is greater than 99.99\%($3.9\sigma$) for all nuclear groups and for all $S$ in the KLEM spectra. The statistical significance could not be calculated with higher precision since there were only 10000 tests in each Monte Carlo run due to a computing limitations. 

Another approach to estimate the statistical significance of the existence of a break in the spectra is to show that the approximation of them by a single power-law function is not adequate. Usually chi-square tests are used to solve similar problems, but here, like in the case of approximation by a double power-law function above, a criterion based on the direct maximum likelihood method for the Poisson statistics should be used. The Poisson criterion uses the amplitude of the likelihood function (\ref{eq:P})  instead of the chi-square value. The distribution of the likelihood function amplitudes has to be modeled by the Monte Carlo method (in contrast to the analytical calculations of the chi-square method). At the output, the criterion gives the probability that the value of the likelihood function does not exceed the value obtained from the experimental spectrum. The large value of the obtained probability indicates that the approximation of the experimental points by a single power-law function is inadequate. The probabilities obtained by this Poisson criterion for experimental spectra are shown in column 4 of Table 2.

It can be seen that, formally, in the IC data, the Poisson criterion does not show that a single power-law spectrum is inadequate. On the contrary, in the KLEM data, the Poisson criterion clearly shows the inadequacy of approximation by a single power-law function for the spectrum of $Z=6\div27$ nuclei and for the spectrum of all particles, and the significance of the inadequacy of a single power-law spectrum for proton and helium spectra is about two and a half standard deviations or higher.

It is not difficult to see that the statistical significance of the break, obtained for the approximation of spectra by double power-law functions, is systematically higher than for single power-law functions. The reason for this is that the method with single power-law functions uses not all available information for the analysis. The method with single power-law functions uses only the fact that there are large deviations of the experimental data from the approximating function, ignoring how these deviations are grouped along the magnetic rigidity axis. On the contrary, the method with approximation by double power-law spectra uses all available information. In our opinion, it is the method with double power-law functions that is more adequate, since it makes use more of available information. It can be noted that both single and double power-law approaches lead to a high statistical significance of the break for the KLEM data. The conclusion that can be drawn is that there is a strong indication of existence of a break near the magnetic rigidity 10\,TV separtely in the spectra of all considered groups of cosmic ray nuclei, therefore this break indicates a universal character.

It is rather difficult to give quantitative estimates of the statistical reliability for the break position $R_*$ due to the complicated shapes of the related probability distributions in the Monte Carlo simulation. We can only conclude that the position of the break is the near 10~TV in different nuclear groups up to a multiplicative factor of the order of the unit.

\section*{Conclusions}

The results of this article show a strong indication (the statistical significance is higher than $3\sigma$) that there exits a universal break near the magnetic rigidity 10\,TV in all nuclear components of cosmic rays. Since the acceleration of cosmic ray particles proceed mainly due to the interaction of particles with magnetic fields in  supernova remnants and expressed in terms of magnetic rigidity of the particles, then the existence of universal knee in terms of magnetic rigidity in the spectra of nuclei is probably connected with the cosmic-ray acceleration limit by some generic or nearby source of cosmic rays. Additional studies are required to refine the parameters of the 10\,TV-knee and to improve the statistical reliability of its existence. 

We are grateful to ROSCOSMOS State Space Corporation and the Russian Academy of Sciences for their continued support of this research.


\vfill

\pagebreak
\onecolumn

\section*{Supplemental Material}

\begin{figure}[h]
\centering
\includegraphics[width=0.48\textwidth,height=0.29\textwidth]{Fig4a.pdf}
\includegraphics[width=0.48\textwidth,height=0.29\textwidth]{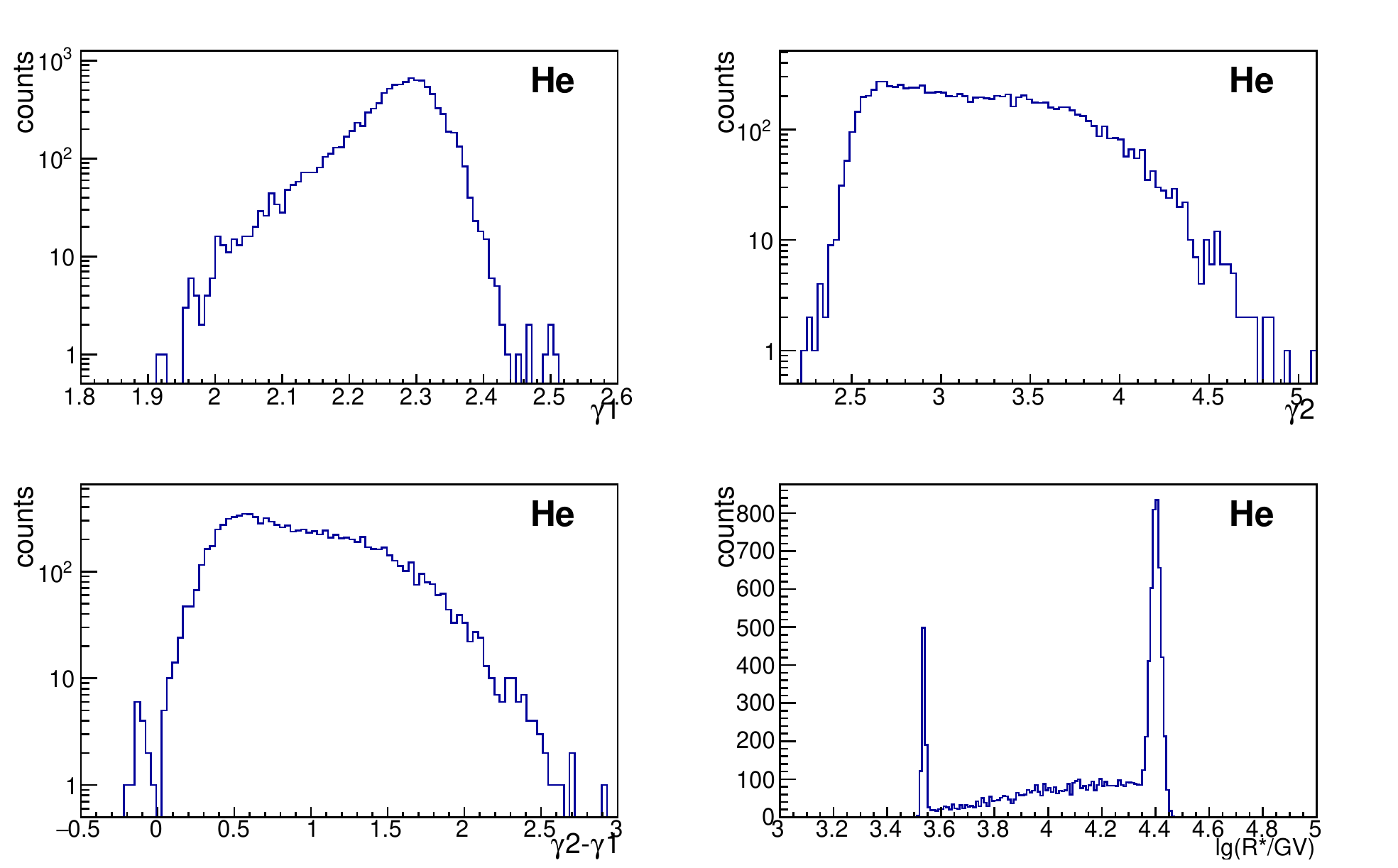}\\
\includegraphics[width=0.48\textwidth,height=0.29\textwidth]{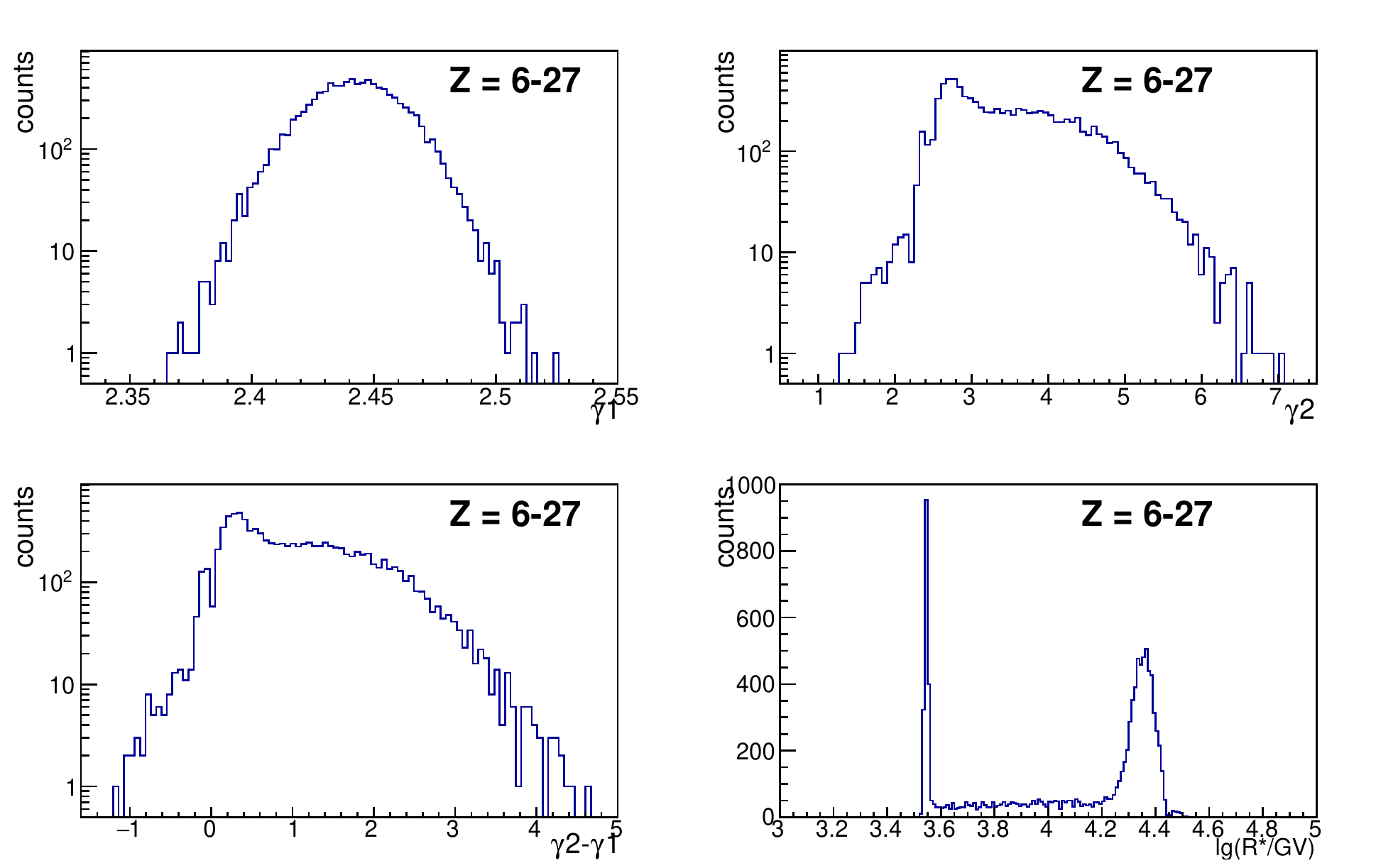}
\includegraphics[width=0.48\textwidth,height=0.29\textwidth]{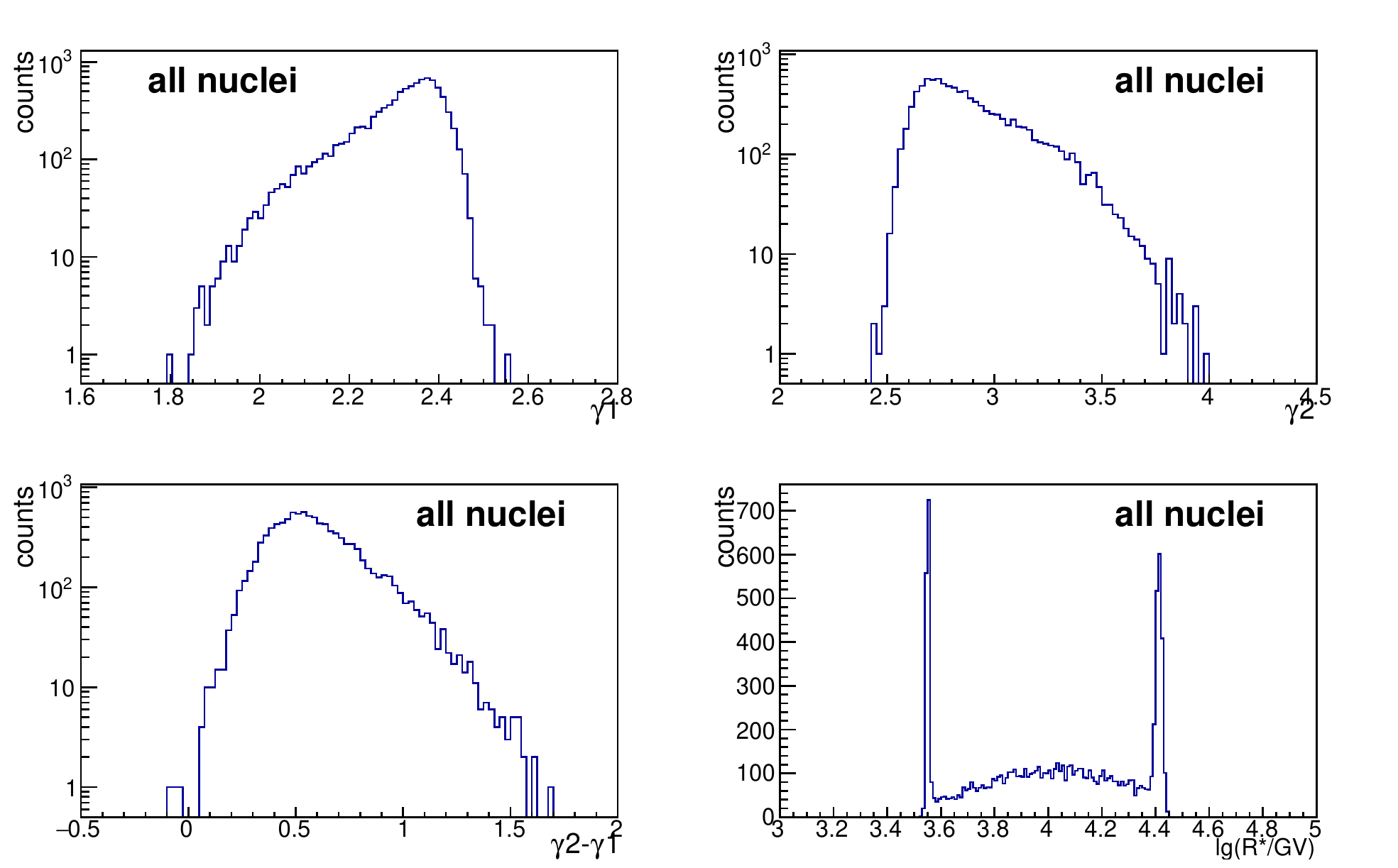}
\caption{The distributions of the values $\gamma_1,\gamma_2,\gamma_2-\gamma_1,R_*$ for IC, smoothing parameter $S=2$}\label{Fig:4} 
\end{figure}

\begin{figure}
\centering
\includegraphics[width=0.48\textwidth,height=0.29\textwidth]{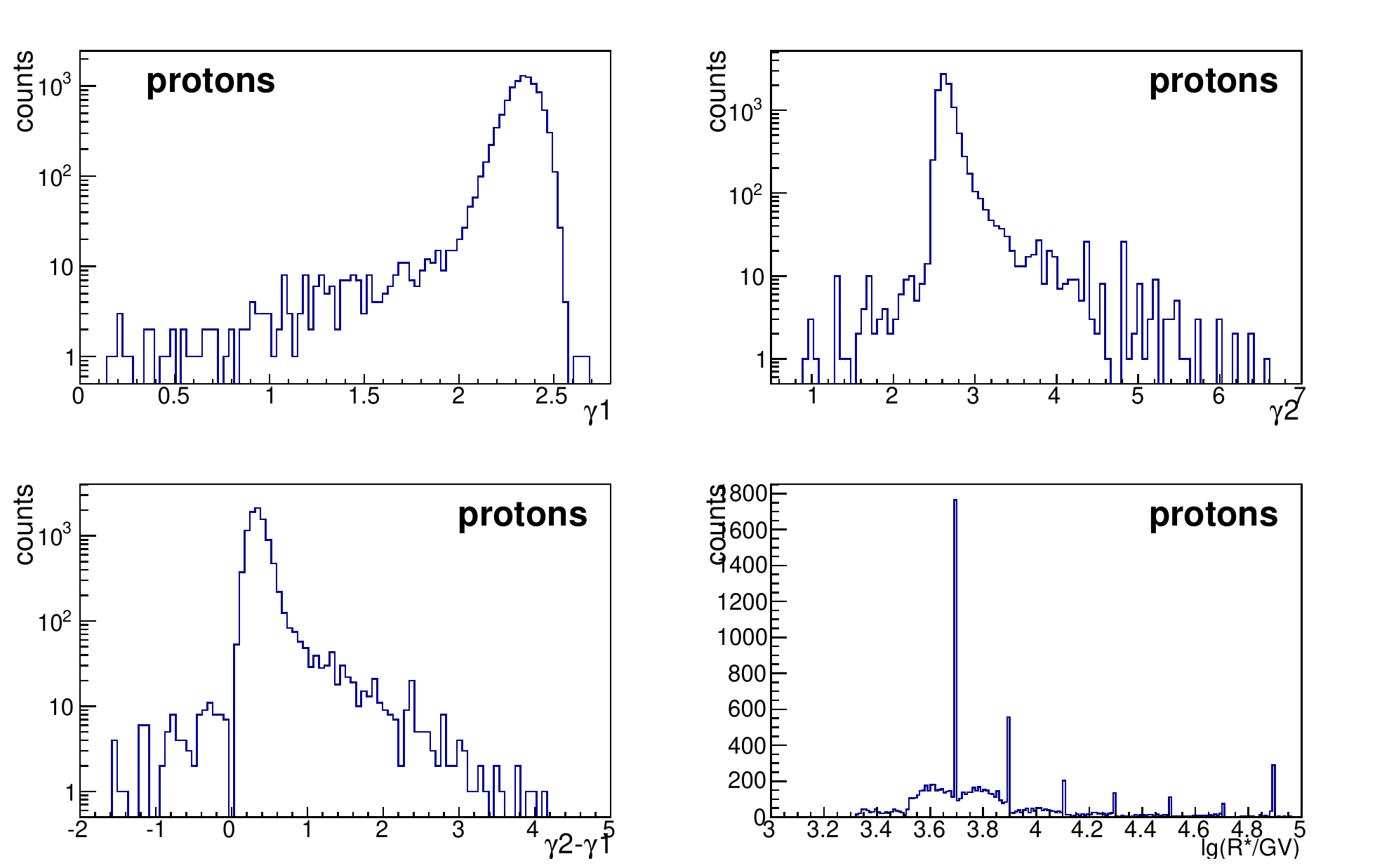}
\includegraphics[width=0.48\textwidth,height=0.29\textwidth]{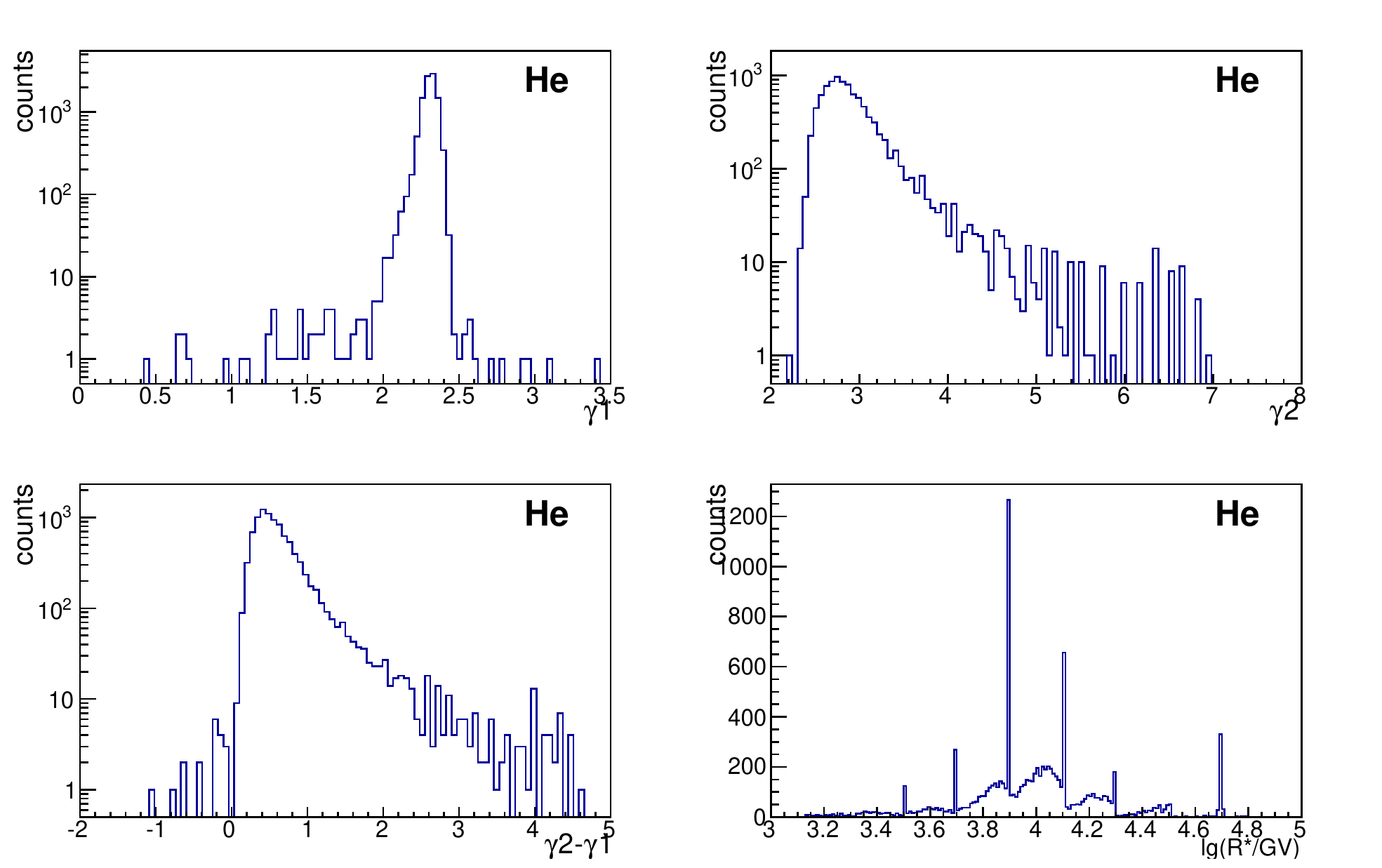}\\
\includegraphics[width=0.48\textwidth,height=0.29\textwidth]{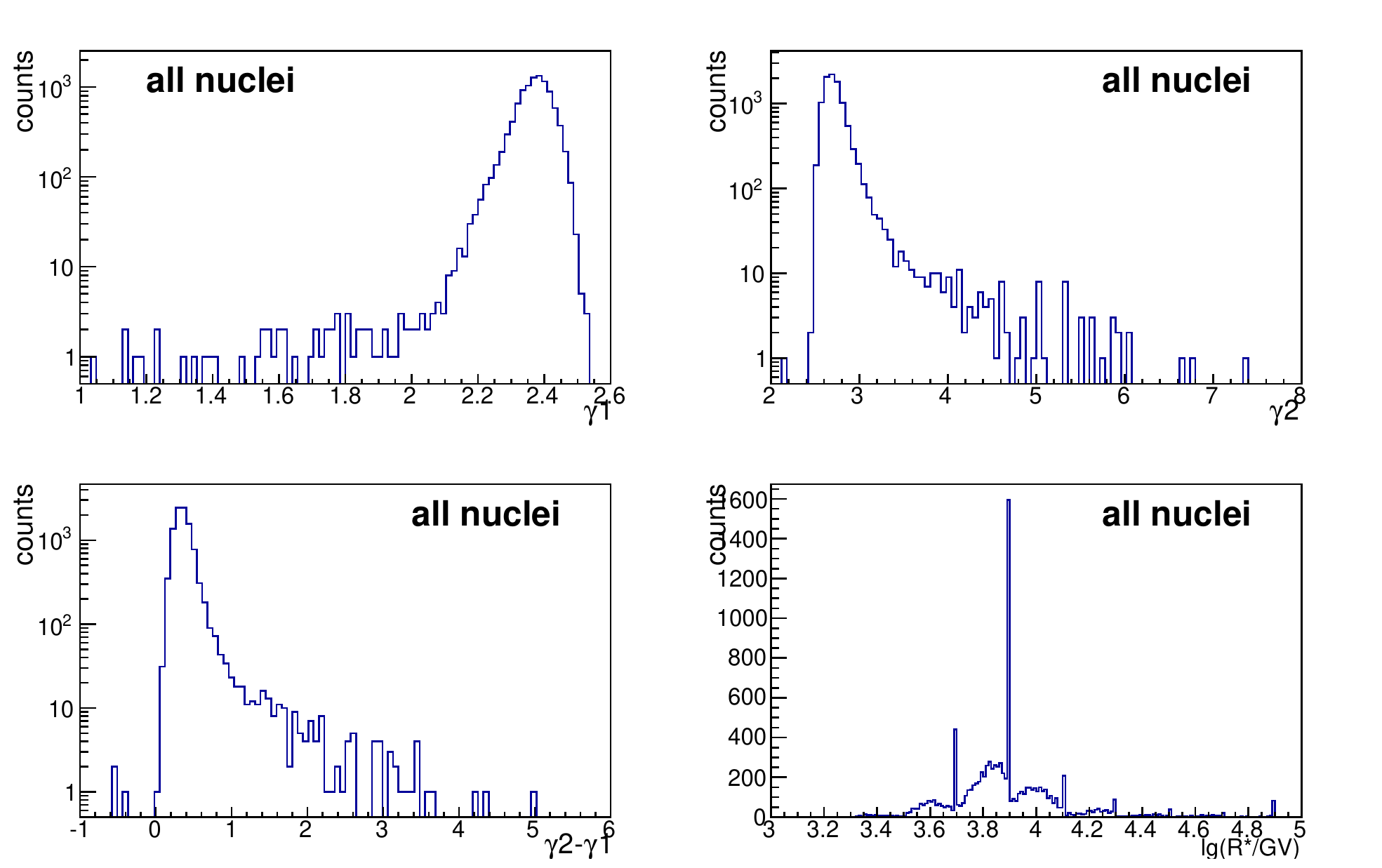}
\includegraphics[width=0.48\textwidth,height=0.29\textwidth]{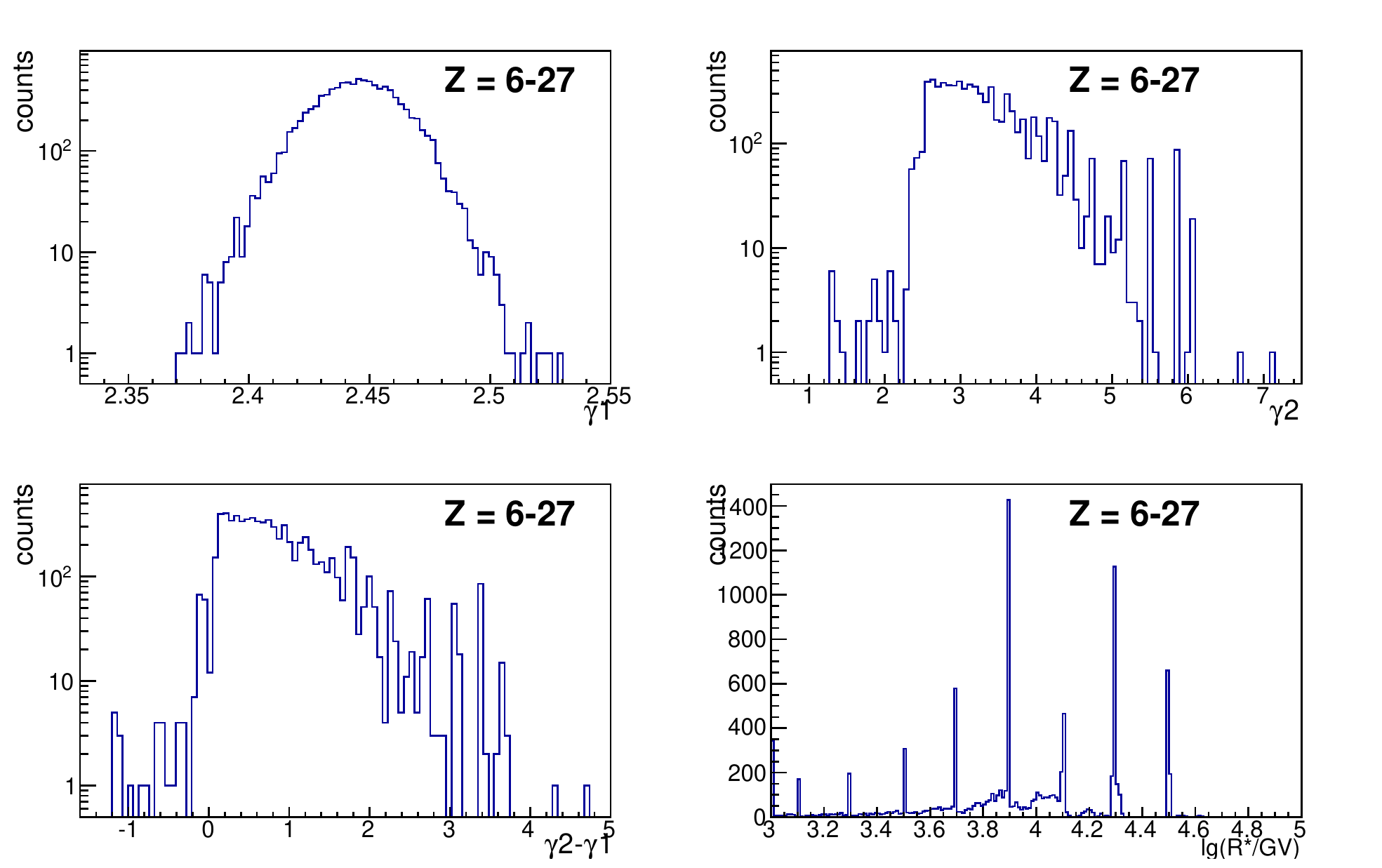}
\caption{The distributions of the values $\gamma_1,\gamma_2,\gamma_2-\gamma_1,R_*$ for IC, smoothing parameter $S=\infty$}\label{Fig:5} 
\end{figure}

\begin{figure}
\centering
\includegraphics[width=0.48\textwidth,height=0.29\textwidth]{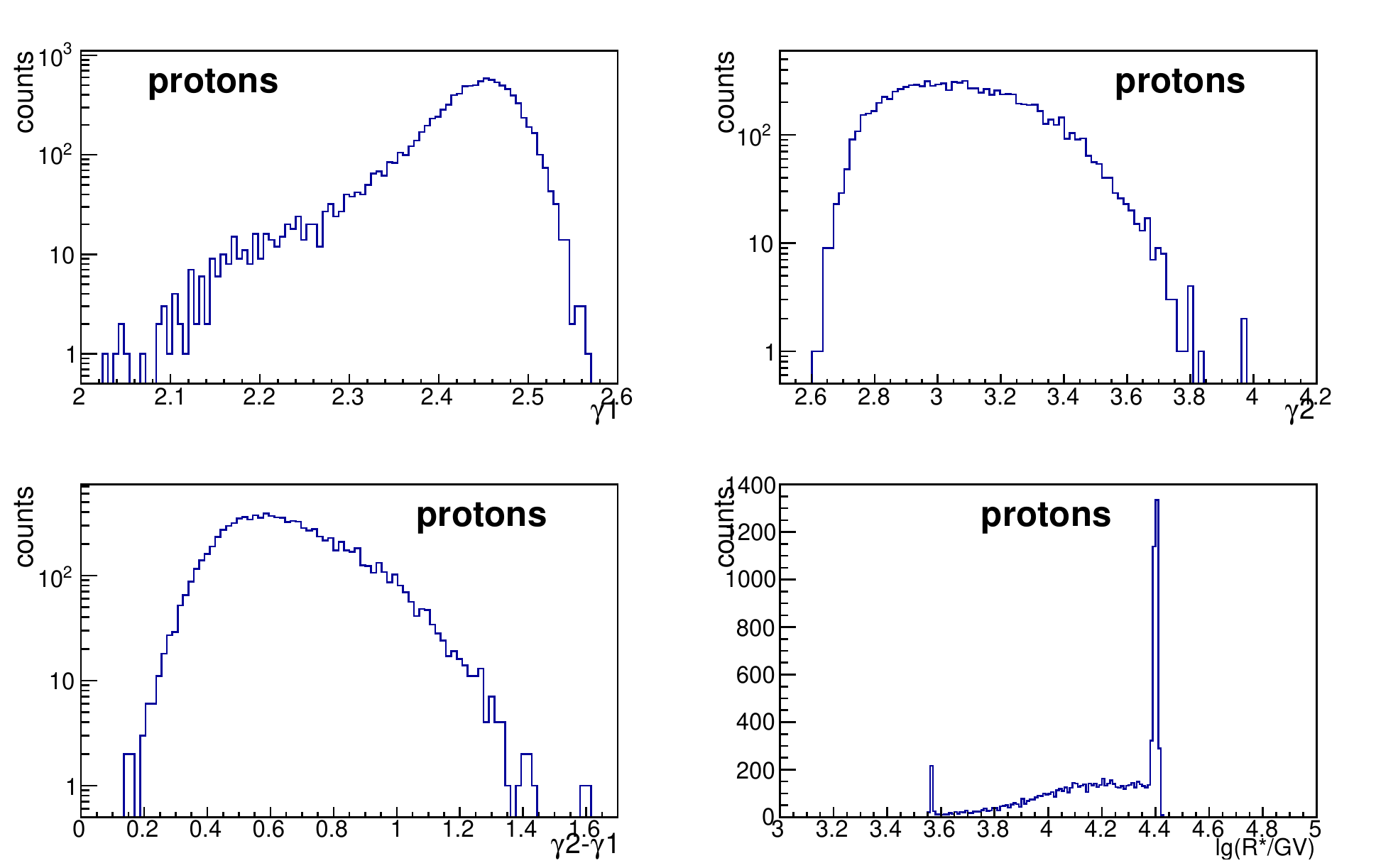}
\includegraphics[width=0.48\textwidth,height=0.29\textwidth]{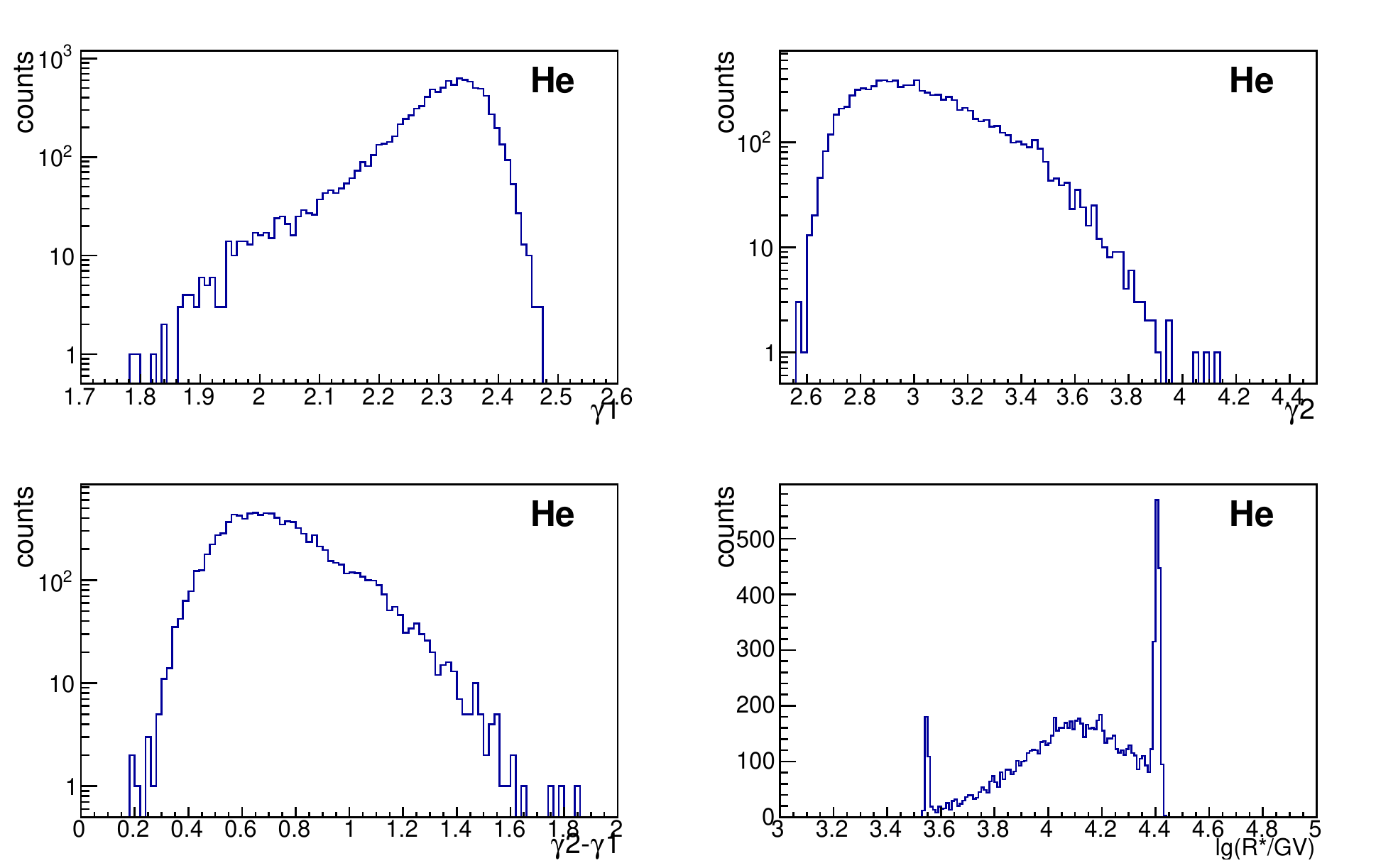}\\
\includegraphics[width=0.48\textwidth,height=0.29\textwidth]{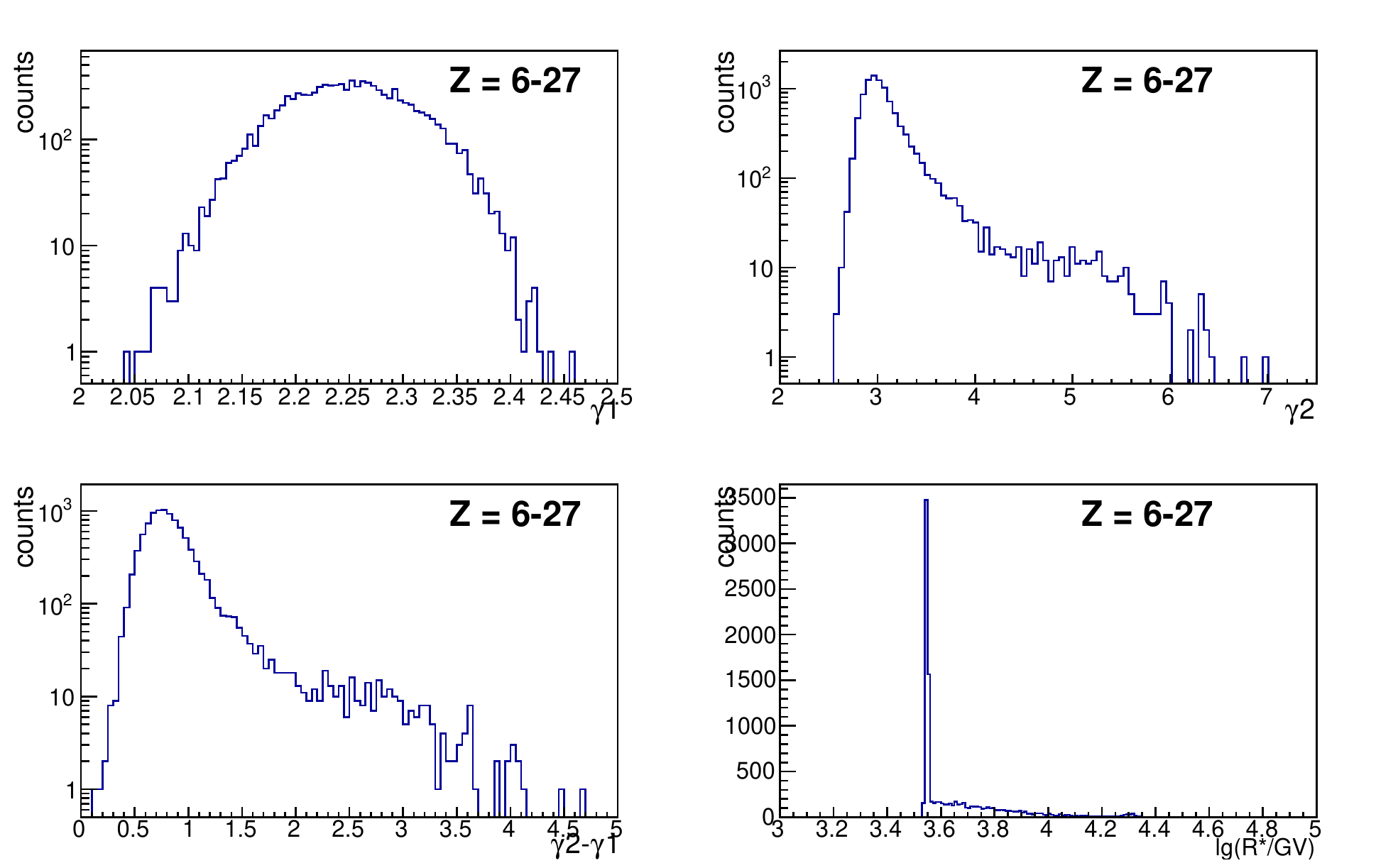}
\includegraphics[width=0.48\textwidth,height=0.29\textwidth]{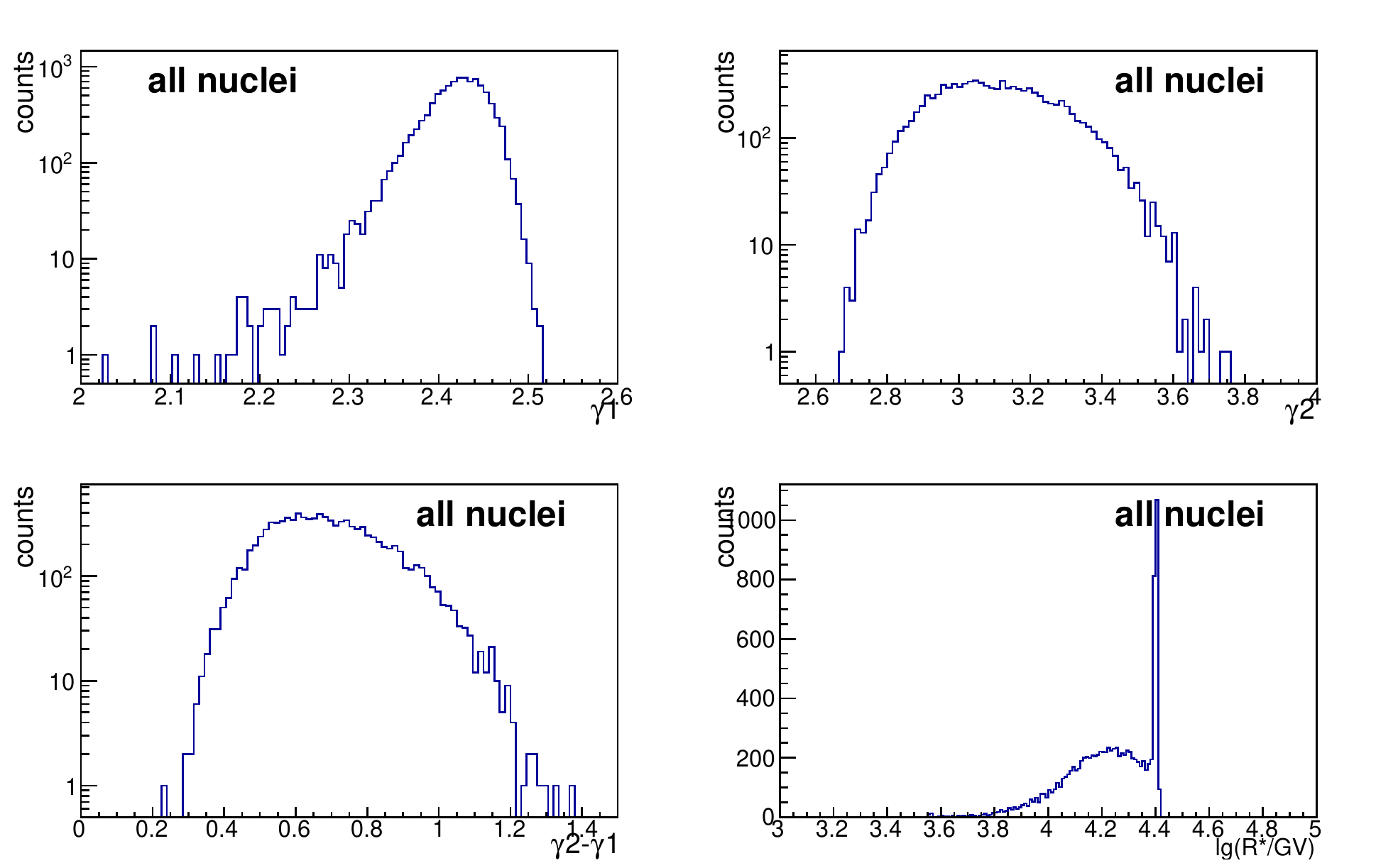}
\caption{The distributions of the values $\gamma_1,\gamma_2,\gamma_2-\gamma_1,R_*$ for KLEM, smoothing parameter $S=2$}\label{Fig:6} 
\end{figure}

\begin{figure}
\centering
\includegraphics[width=0.48\textwidth,height=0.29\textwidth]{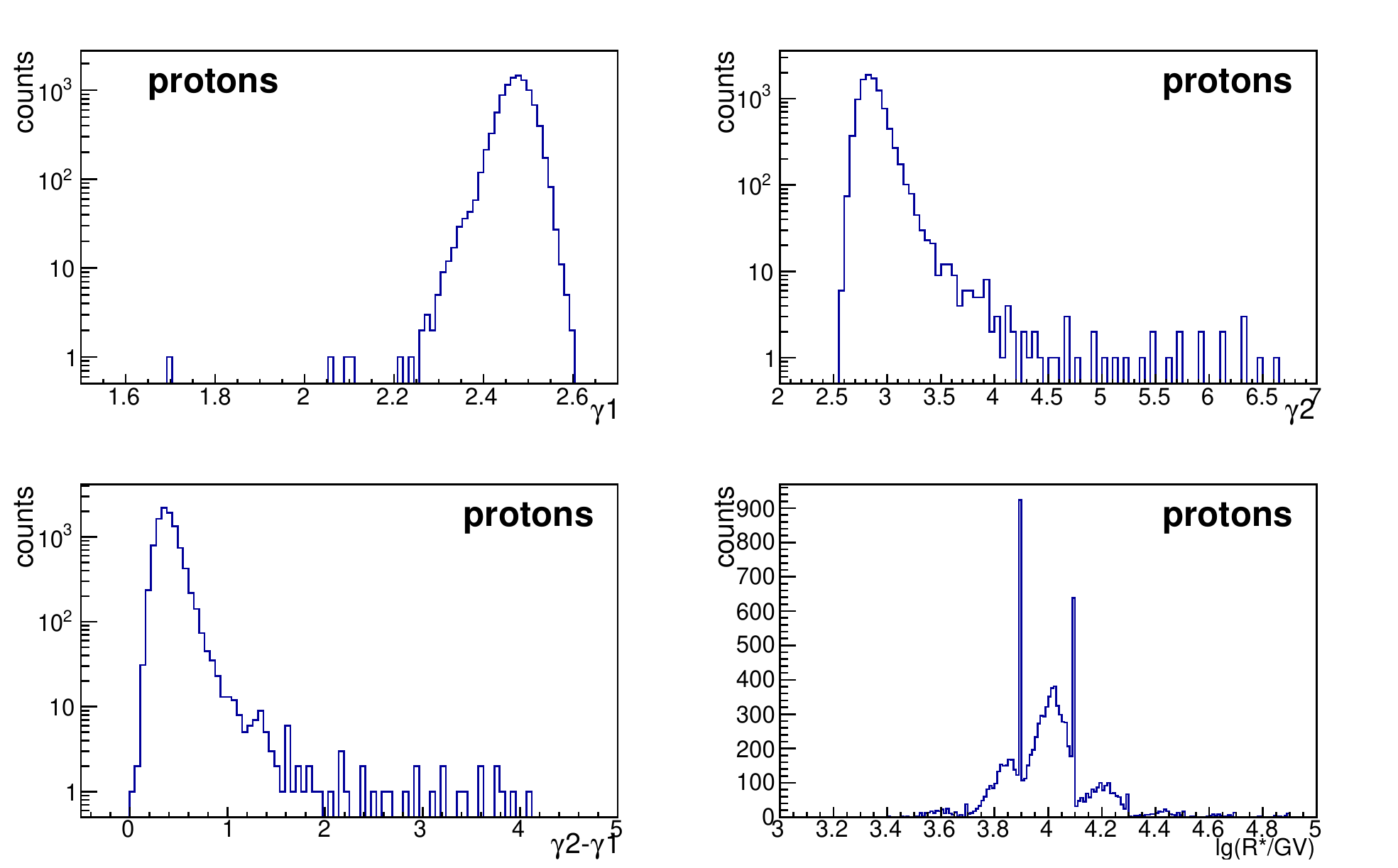}
\includegraphics[width=0.48\textwidth,height=0.29\textwidth]{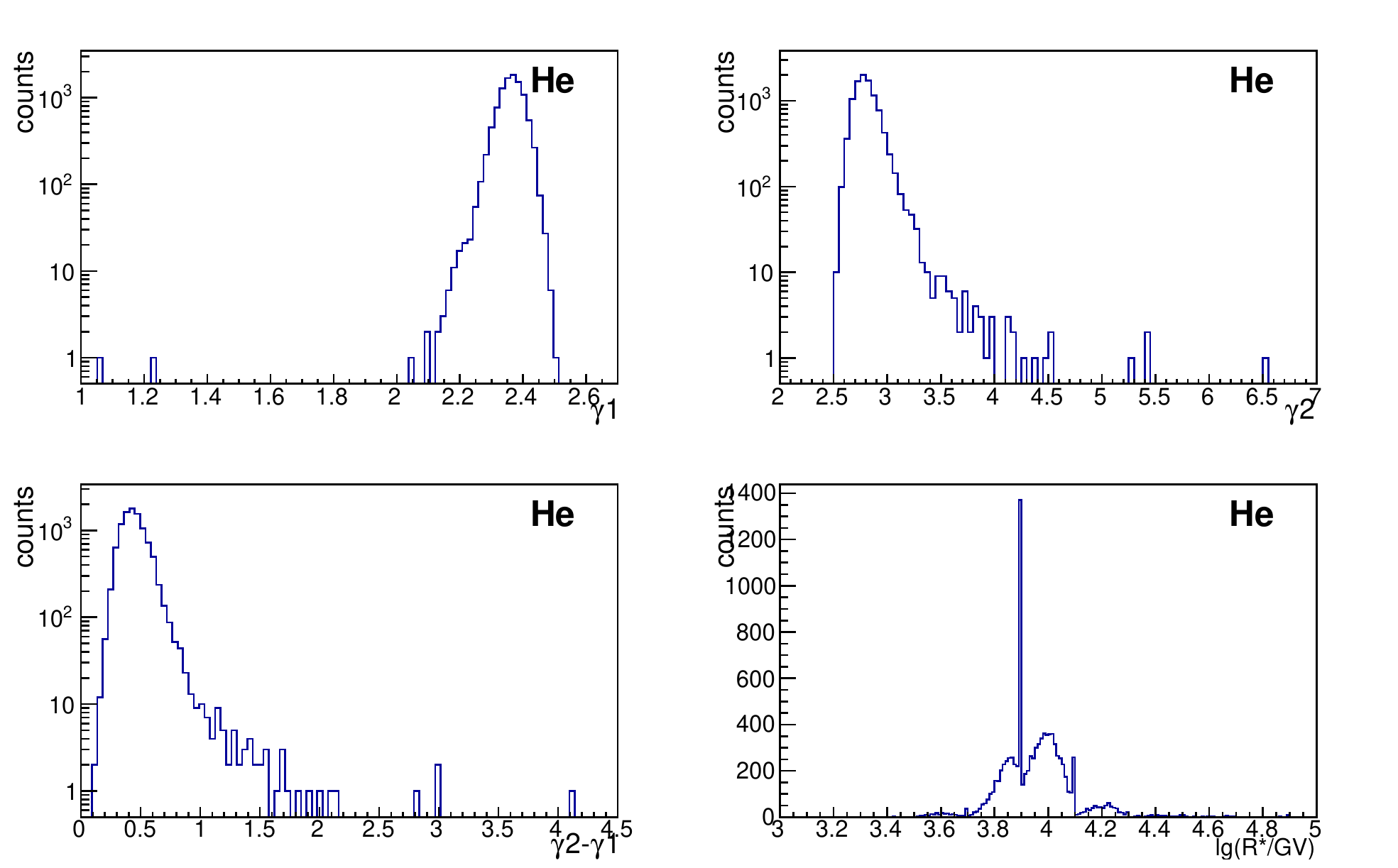}\\
\includegraphics[width=0.48\textwidth,height=0.29\textwidth]{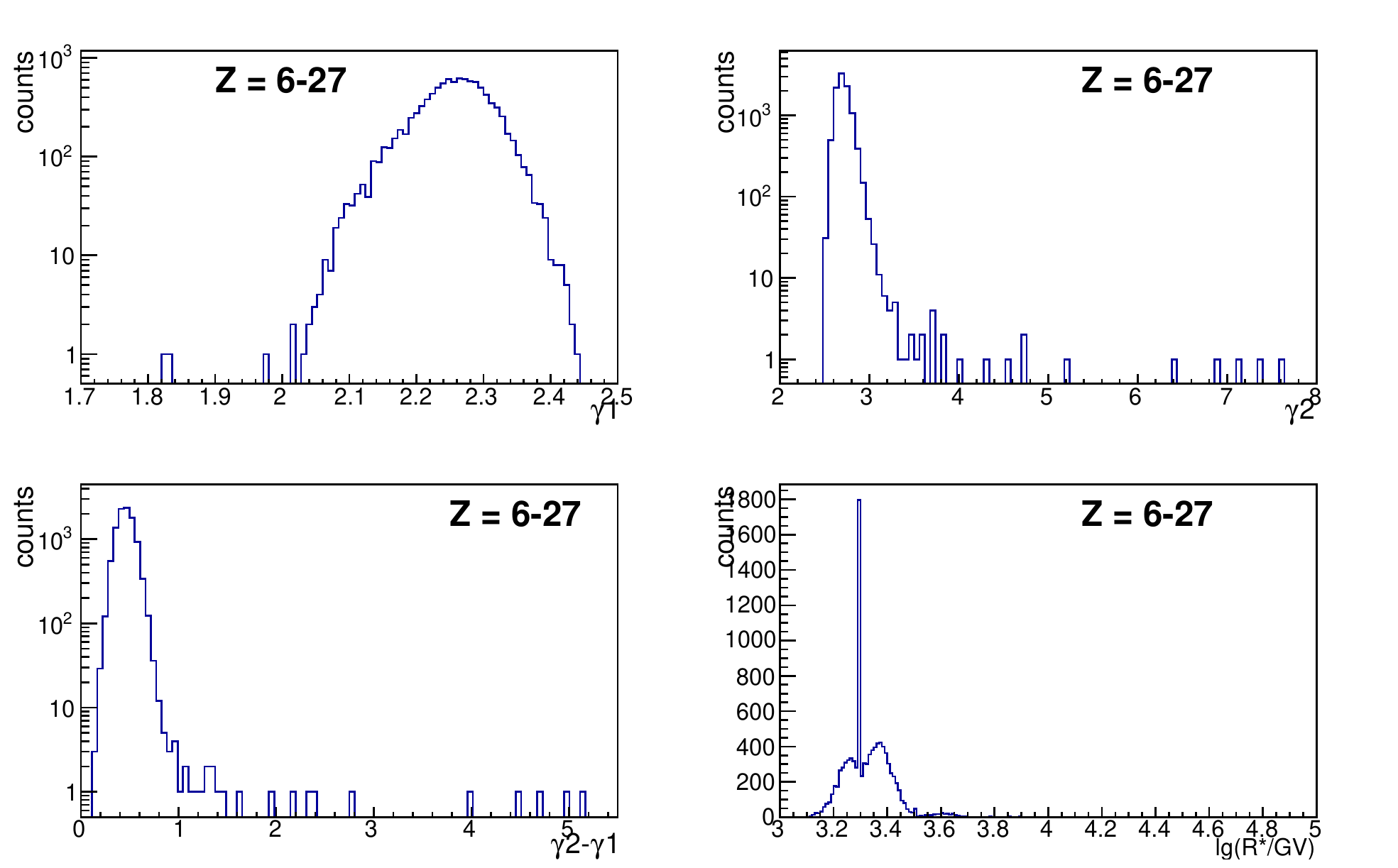}
\includegraphics[width=0.48\textwidth,height=0.29\textwidth]{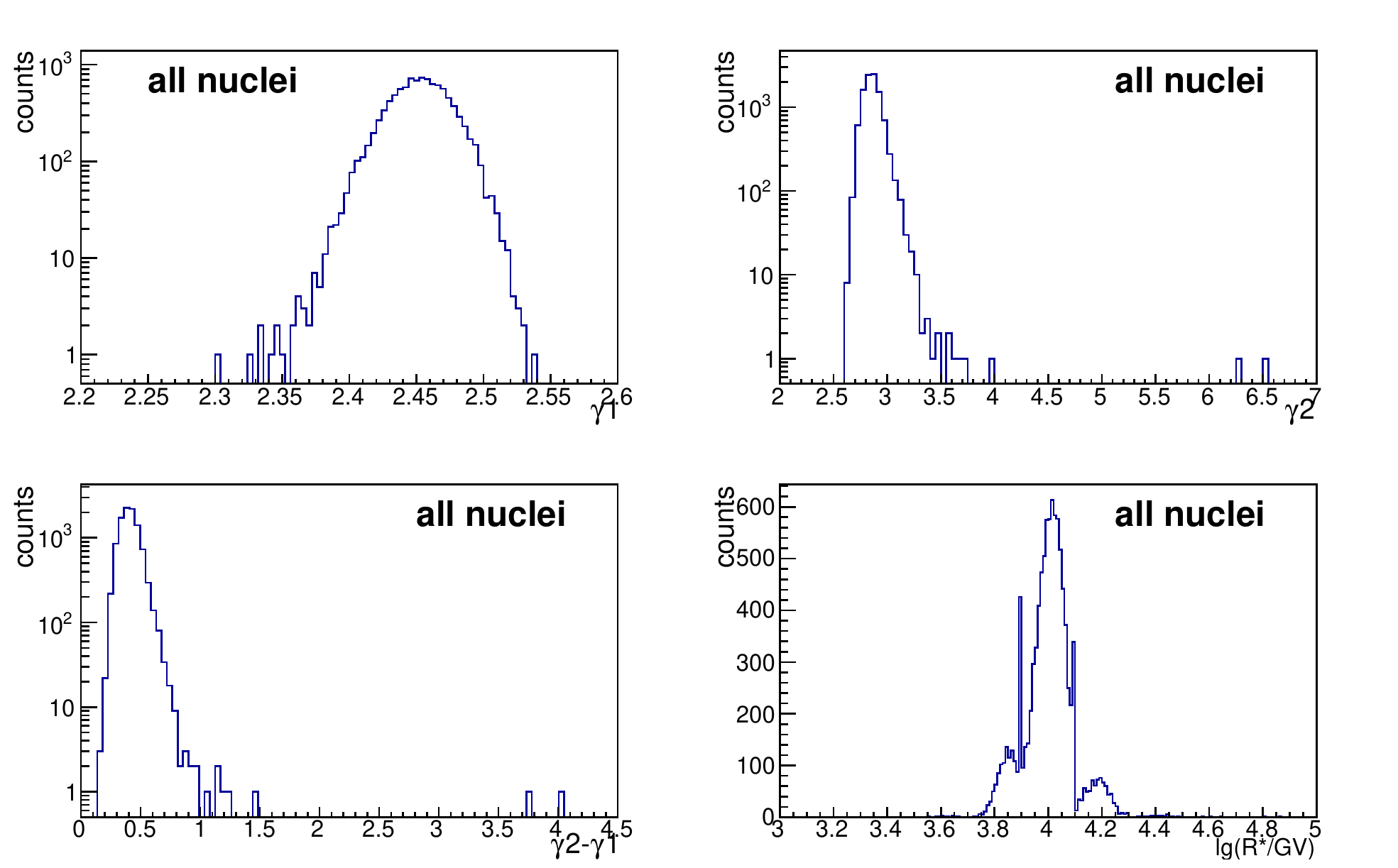}
\caption{The distributions of the values $\gamma_1,\gamma_2,\gamma_2-\gamma_1,R_*$ for KLEM, smoothing parameter $S=\infty$}\label{Fig:7} 
\end{figure}

In Fig.~\ref{Fig:4}--Fig.~\ref{Fig:7} are shown the distributions of the values $\gamma_1,\gamma_2,\gamma_2-\gamma_1,R_*$ that were obtained in the Monte Carlo simulation (10000 test in each run) as described in the paper for all of the studied nuclear groups, for smoothing parameters $S=2$ and $S=\infty$, for both methods of energy determination IC and KLEM. 


\end{document}